\documentclass[fleqn,usenatbib]{mnras}

\usepackage{subfig}

\usepackage[T1]{fontenc}

\DeclareRobustCommand{\VAN}[3]{#2}
\let\VANthebibliography\thebibliography
\def\thebibliography{\DeclareRobustCommand{\VAN}[3]{##3}\VANthebibliography}

\usepackage{graphicx}	
\usepackage{amsmath}	
\usepackage{amssymb}	
\usepackage{ulem}       
\usepackage{lineno}

\title[Multi-band optical variability of blazars]{Multi-band optical variability of a newly discovered twelve blazars sample from 2013 - 2019}

\author[M. D. Jovanovi\'c et al.]{
	Miljana D. Jovanovi\'c,$^{1,2}$\thanks{E-mail: miljana@aob.rs (MDJ)}
	Goran Damljanovi\'c,$^{1}$
	Fran\c{c}ois Taris,$^{3}$
	Alok C. Gupta,$^{4,5}$
	and Gopal Bhatta$^{6}$
	\\
	\\
	$^{1}$Astronomical observatory, Volgina 7, 11060 Belgrade, Serbia\\
	$^{2}$Department of Astronomy, Faculty of Mathematics, University of Belgrade, Studentski trg 16, 11000 Belgrade, Serbia\\
	$^{3}$Observatoire de Paris - SYRTE, 61 av. de l'Observatoire, 75014 Paris, France\\
	$^{4}$Key Laboratory for Research in Galaxies and Cosmology, Shanghai Astronomical Observatory, Chinese Academy of Sciences, \\
~Shanghai 200030, China\\
	$^{5}$Aryabhatta Research Institute of Observational Sciences (ARIES), Manora Peak, Nainital 263001, India\\
	$^{6}$Institute of Nuclear Physics, Polish Academy of Sciences PAN PL--31342 Krakow, Poland\\
}

\date{Accepted XXX. Received YYY; in original form ZZZ}

\pubyear{2015}

\begin{document}
	\label{firstpage}
	\pagerange{\pageref{firstpage}--\pageref{lastpage}}
	\maketitle
	
	\begin{abstract}
		Here we present the first optical photometric monitoring results of a sample of twelve newly discovered blazars from the ICRF – Gaia CRF astrometric link. The observations were performed from April 2013 until August 2019 using eight telescopes located in Europe.  For a robust test for the brightness and colour variability, we use Abbé criterion and F–test. Moreover, linear fittings are performed to investigate the relation in the colour-magnitude variations of the blazars. Variability was confirmed in the case of 10 sources; two sources, 1429+249 and 1556+335 seem to be possibly variable. Three sources (1034+574, 1722+119, and 1741+597) have displayed large amplitude brightness change of more than one magnitude. We found that the seven sources displayed bluer-when-brighter variations, and one source showed redder-when-brighter variations. We briefly explain the various AGN emission models which can explain our results.
	\end{abstract}
	
	
	
	
	\section{Introduction}
	
	Blazars form a subclass of radio loud (RL) active galactic nuclei (AGN) which eject relativistic jets along the observer's line of sight \citep{1995PASP..107..803U}. BL Lacertae objects (BL Lacs) and flat spectrum radio quasars (FSRQs) are collectively referred to as blazars. In the composite optical/UV spectra, BL Lacs show featureless continuum or weak narrow emission lines (equivalent width EW $\leq$ 5\AA) \citep[e.g.,][]{1991ApJ...374..431S,1996MNRAS.281..425M} while FSRQs show prominent broad emission lines. Blazars flux, polarization and spectra are highly variable in the whole (radio to $\gamma-$rays) electromagnetic (EM) spectrum \citep[e.g.,][and references therein]{2017MNRAS.472..788G}. Blazars in general show variability on diverse timescales, ranging as short as a few minutes to as long as several decades. Variability timescales of blazars can be broadly divided into three classes: timescale from a few minutes to a less than a day is commonly known as microvariability \citep{1989Natur.337..627M} or intra-day variability (IDV) \citep{1995ARA&A..33..163W} or intra-night variability \citep{1993MNRAS.262..963G}. Variability timescales ranging from a few days to a few months are called short-term variability (STV), and those from a few months to several years are termed as long-term variability \citep[LTV: see][]{2004A&A...422..505G}. \\
	\\ 
	Blazar emission extending across the entire electromagnetic (EM) spectrum is dominated by nonthermal radiation from the relativistic jets. The broadband emission provides us an excellent opportunity to study their spectral energy distribution (SED), often characterised by well-known double-hump structure \citep{1995ApJ...440..525V,1998MNRAS.299..433F}. In all classes of blazars, from radio to soft X-ray frequencies the dominant emission mechanism is synchrotron emission, and whereas in hard X-ray and $\gamma-$ray energies it is most probably inverse Compton (IC) scattering \citep{1997ARA&A..35..445U,2007Ap&SS.309...95B}. 
	In the recent classification scheme based on the peak synchrotron frequency $\nu_{peak}$, blazars are classified in three sub-classes: LSP - Low synchrotron peak with $\nu_{peak} \leq {\rm10^{14}}$ Hz, ISP - Intermediate synchrotron peak with ${\rm10^{14}} < \nu_{peak} < {\rm10^{15}}$ Hz, and HSP - High synchrotron peak with $\nu_{peak} \geq {\rm10^{15}}$ Hz \citep{2010ApJ...716...30A}. \\
	
	\begin{table*}
		\caption{Optical Observation Log for the sample blazars}  
		\label{table:1} 
		\centering     
		\begin{tabular}{c c c c c r l c}  
			\hline\hline   
			IERS name & $\alpha_{J2000.0}$($\degr$) & $\delta_{J2000.0}$($\degr$) & $z$ & AGN & \multicolumn{2}{c}{Observation duration} & No.~of obs. \\
			& & & & Type & dd mm yyyy & dd mm yyyy & ~$V$, ~$R$ \\\hline
			0049+003 & ~~13.02321 & 0.593930 & 0.399714 & FSRQ & 06 09 2013 & 08 08 2019 & 30, 40 \\
			0907+336 & 137.65431 & 33.49012 & 0.354000 & BL Lac & 14 04 2013 & 06 04 2019 & 39, 42 \\
			1034+574 & 159.43461 & 57.19878 & 1.095700 & BL Lac & 09 07 2013 & 07 04 2019 & 47, 47 \\
			1212+467 & 183.79143 & 46.45420 & 0.720154 & FSRQ & 09 07 2013 & 31 03 2019 & 50, 50 \\
			1242+574 & 191.29167 & 57.16510 & 0.998229 & BL Lac & 02 04 2014 & 06 08 2019 & 49, 57 \\
			1429+249 & 217.85787 & 24.70575 & 0.406590 & BL Lac / FSRQ & 04 04 2014 & 06 08 2019 & 40, 44 \\
			1535+231 & 234.31043 & 23.01127 & 0.462515 & BL Lac / FSRQ & 04 04 2014 & 06 08 2019 & 43, 44 \\
			1556+335 & 239.72993 & 33.38850 & 1.653598 & FSRQ & 04 04 2014 & 06 08 2019 & 41, 50 \\
			1607+604 & 242.08560 & 60.30784 & 0.178000 & BL Lac & 08 07 2013 & 06 08 2019 & 42, 48 \\
			1612+378 & 243.69564 & 37.76869 & 1.531239 & FSRQ & 09 07 2013 & 06 08 2019 & 37, 42 \\
			1722+119 & 261.26810 & 11.87096 & 0.340000 & BL Lac & 09 07 2013 & 08 08 2019 & 43, 47 \\
			1741+597 & 265.63334 & 59.75186 & 0.415000 & BL Lac & 09 07 2013 & 07 08 2019 & 55, 62 \\\hline      
		\end{tabular}
	\end{table*}
	
	\noindent
	The optical band is quite narrow in comparison to the other spectral bands over the entire EM spectrum. Nevertheless it helps us obtain important information regarding nonthermal synchrotron emission as well as possible thermal emission from accretion disc. In general on STV and LTV timescales, spectral trends of bluer-when-brighter (BWB) in BL Lacs and redder-when-brighter (RWB) in FSRQs have been observed, although occasionally opposite trends are also detected in some blazars \citep[e.g.,][and references therein]{2006A&A...450...39G,2012MNRAS.425.3002G,2017ApJ...844..107I}. In recent times, extensive studies on blazar optical variability on diverse timescales have been carried out using observations from both space and ground based telescopes. The results demonstrate that the general nature of the LTV is mostly characterised by substantial change in the flux, which are occasionally accompanied by sudden flares and quasi-periodic oscillations \citep{2023MNRAS.tmp..320B}. The blazars light curves often show normal or log-normal flux distribution \citep[see][]{2021ApJ...923....7B}. In shorter timescales similar variability properties with power-law spectral density were reported in a large number of TESS blazar light curves \citep{2023MNRAS.518.1459P}. \\
	\\
	On June 13, 2022 the third data release (DR3) of Gaia mission was made available for public \citep{refId0}. The Gaia uses astrometric observations of optical counterparts of sources from the radio catalogue International Celestial Reference Frame (ICRF) \citep{2020A&A...644A.159C} to adjust its reference frame. A set of $\sim$ 1.6 million quasi-stellar objects (QSOs) constitutes the third version of the Gaia celestial reference frame (Gaia-CRF3). 398 sources not included in the ICRF list were mentioned as potential sources for VLBI (Very Long Baseline Interferometry) observations \citep{2010A&A...520A.113B}. From this list, 105 sources were observed with a global VLBI array which detected 47 point-like sources on VLBI scales and classified as AGNs \citep{2011A&A...526A.102B}.\\
	\\
	From 2013 to 2019, we conducted optical photometric observations in the $V$ and $R$ bands for 12 blazars selected from a sample of 47 AGNs detected in \citet{2011A&A...526A.102B}. Of the 12 blazars studied, 6 are classified as BL Lacs, 4 as FSRQs, and 2 exhibit characteristics of both BL Lacs and FSRQs. The detailed information about these blazars: their International Earth Rotation Service (IERS) name and observations log are provided in Table \ref{table:1}. In this work, we conduct a thorough investigation of the optical flux and colour variability properties of these blazars on both short-term and long-term timescales. \\
	\\	
	The paper is structured as follows. In Section \ref{sec:2}, we describe our new photometric observations. The detailed description of various analysis techniques used is explained in Section \ref{sec:3}. Section \ref{sec:4} gives the results of individual AGN. Discussion and Conclusions are given in Section \ref{sec:5}.
	
	\section{Observations and Photometry}\label{sec:2}
	
	The optical photometric observations of the blazars were performed using eight telescopes located in Europe. Out of these eight telescopes, two are stationed at Astronomical Station Vidojevica (ASV) of Astronomical Observatory of Belgrade, Serbia; one robotic Joan Or\'o telescope (TJO) at the Montsec Astronomical Observatory, Catalonia, Spain; four telescopes in Bulgaria of which three at Rozhen, NAO and one in Belogradchik; and one telescope at Leopold Figl at Vienna, Austria. The details about these telescopes, their mirror aperture, mounted CCD cameras and optical filters are presented in Table \ref{table:telescopes}. \\
	\\
	During each observing night, two or more CCD image frames of the blazars were acquired in both the $V$ and $R$ bands. The image processing was performed using IRAF\footnote{Image Reduction and Analysis Facility, a general purpose software system for the reduction and analysis of astronomical data. IRAF is distributed by the National Optical Astronomy Observatories, which are operated by the Association of Universities for Research in Astronomy, Inc., under cooperative agreement with the National Science Foundation.} scripting language (ascl:9911.002) \citep{1986SPIE..627..733T,1993ASPC...52..173T}. Bias, dark, and flat-field frames were obtained for every observing night, which were used for advanced image calibration and bad pixel mapping (dark frames for hot, and flat-field for dead pixel map). In addition, the corrections for cosmic rays were performed using Laplacian Cosmic Ray Identification method \citep{2001PASP..113.1420V}. \\
	\\	
	We performed differential photometry using Maxim DL software for determining the brightness of the sources with the aperture radius of $\sim$6 arcsec. The details about differential photometry and selection of comparison and control stars are presented in papers \citet{2018A&A...611A..52T}, and \citet{2019SerAJ.199...55J}. The PSF $ugriz$ (point spread function $u$, $g$, $r$, $i$, and $z$) magnitudes for the comparison and control stars were taken from the Sloan Digital Sky Survey Data Release 14 (SDSS DR14) catalogue \citep{2018ApJS..235...42A}. The magnitudes in $V$, and $R$ bands were calculated from $g$, $r$, and $i$ band magnitudes using the equations given by \citet{2008AJ....135..264C}. Except for source 1722+119 magnitudes $V$, and $R$ are taken from paper \citet{2014Ap.....57..176D}.
	
	\begin{table*}
		\caption{Details of telescopes and instruments}  
		\label{table:telescopes} 
		\centering     
		\begin{tabular}{c c c c c }  
			\hline\hline   
			Telescope & ASV 60cm & ASV 1.4m & TJO 80cm & Rozhen 2m \\\hline
			CCD Model & Apogee Alta E47 & Andor iKon-L & Andor iKon-L & Andor iKon-L \\
			& Apogee Alta U42 & Apogee Alta U42 & FLI PL4240-1-B & VersArray:1300B \\
			& SBIG ST10 XME &   &  & \\
			Chip Size (pixels) & 1024$\times$1024 & 2048$\times$2048 & 2048$\times$2048 & 2048$\times$2048 \\
			& 2048$\times$2048 & 2048$\times$2048 & 2048$\times$2048 & 1340$\times$1300 \\
			& 2184$\times$1472 &   &   & \\
			Scale (arcsec/pixel) & 0.45  & 0.244  & 0.361  & 0.176 \\
			& 0.466  & 0.244  & 0.364  & 0.258 \\
			& 0.23  &   &   & \\
			Field (arcmin$^{2}$) & 7.6$\times$7.6 & 8.3$\times$8.3 & 12.3$\times$12.3 & 6.0$\times$6.0 \\
			& 15.8$\times$15.8 & 8.3$\times$8.3 & 12.3$\times$12.3 & 5.76$\times$5.76 \\
			& 8.4$\times$7.5 &  &   & \\
			Gain ($\rm{e}^{-}$/ADU) & 2.56  & 1  & 1  & 1.7 \\
			& 1.25  & 1.25  & 1.5  & 1 \\
			& 1.2  &  &   & \\
			Read-out Noise ($\rm{e}^{-}$ rms) & 37.2 & 7  & 13.7  & 6 \\
			& 12.5 & 12.5  & 6.6  & 2 \\
			& 8.8 &  &   & \\
			Typical Seeing (arcsec)  & 1-2 & 1-2 & 1-2  & 1.5-2.5 \\\hline
			Telescope & LFOA 1.5m & Rozhen 50/70cm & Rozhen 60cm & Belogradchik 60cm \\\hline
			CCD Model & SBIG ST10 XME & FLI PL16803 & FLI PL9000 & FLI PL9000 \\
			Chip Size (pixels) & 2184$\times$1472 & 4096$\times$4096 & 3056$\times$3056 & 3056$\times$3056 \\
			Scale (arcsec/pixel) & 0.15 & 1.079 & 0.33 & 0.33 \\
			Field (arcmin$^{2}$) & 5.6$\times$3.8 & 73.66$\times$73.66 & 16.8$\times$16.8 & 16.8$\times$16.8 \\
			Gain ($\rm{e}^{-}$/ADU) & 1.42 & 1 & 1 & 1 \\
			Read-out Noise ($\rm{e}^{-}$ rms) & 13.38 & 9 & 9 & 9 \\
			Typical Seeing (arcsec) & 2-4 & 2-4 & 1.5-2.5 & 1.5-2.5 \\\hline
			
		\end{tabular}
	\end{table*}

	\begin{table}
		\caption{Examples of observations from 2013 to 2019 in $V$, and $R$ bands.}  
		\label{table:example} 
		\centering     
		\begin{tabular}{c c c c c }  
			\hline\hline    
			Name & Julian Date (JD) & Magnitude & Error & band \\\hline   
			0049+003 & 2456542.47938 & 16.296 & 0.021 & $V$ \\
			0049+003 & 2456542.49410 & 15.877 & 0.014 & $R$\\
			0049+003 & 2457011.29274 & 15.849 & 0.009 & $R$\\
			0049+003 & 2457011.29414 & 16.179 & 0.012 & $V$\\
			0049+003 & 2457011.34780 & 15.947 & 0.056 & $R$\\
			0049+003 & 2457011.34970 & 16.303 & 0.005 & $V$\\\hline 
			\multicolumn{5}{l}{Notes. This photometric data table is available in its entirety in a}\\
			\multicolumn{5}{l}{machine-readable form in the online journal. A portion is shown}\\
			\multicolumn{5}{l}{here for guidance regarding its form and content.}
		\end{tabular}
	\end{table}
	
	\begin{figure}
		\centering
		\includegraphics[width=\columnwidth]{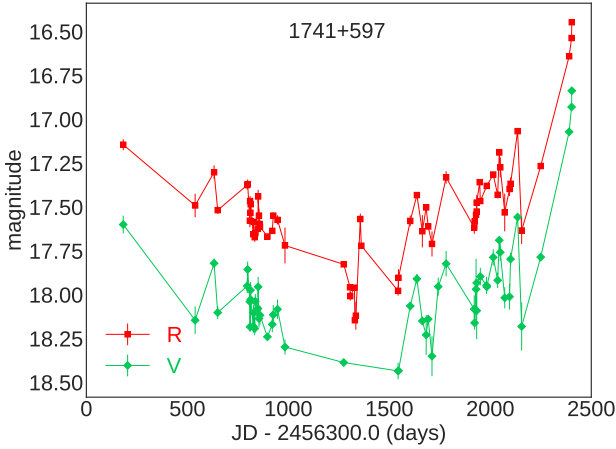}
		\caption{$V$ (green diamonds) and $R$ (red squares) band light curves of 1741+597 from July 2013 to August 2019.
		}
		\label{LC_example}
	\end{figure}

	\section{Analysis methods}\label{sec:3}
	\noindent To test for the presence of variability in the source, we performed two statistics: Abb\'{e}'s criterion and F--test. Both of the tests require normal distribution of data (in some cases the Abb\'{e}' criterion can be applied when data distribution is different from normal \citep{2006Lemeshko}). We consider a light curve of the source as variable if the variability is detected by both test statistics. Before applying these statistical tests, we used 3--$\sigma$ rule \citep{pukelsheim1994three} and Shapiro-Wilk test of normality \citep{razali2011power}. We discarded some of the data which were obtained under poor weather conditions. We concluded that the statistical methods which require normal distribution of data can be applied.
	
	\subsection{Abb\'{e}'s criterion}\label{subsec3.1}
	\noindent
	We used Abb\'{e}'s criterion to determine whether the elements of this sample are stochastically independent or not. Abb\'{e}'s criterion is intended for checking hypotheses that all the observed quantities in the sample have identical mathematical expectations. The criterion is often used for checking the absence of systematic changes in a series of measurements. Abb\'{e}'s statistic $q$ is defined as the ratio of the Allan variance $\sigma_{AV}$ and unbiased sample variance $\sigma_{D}$
	\begin{equation}
	q = \frac{\sigma_{AV}}{\sigma_{D}}=\frac{\dfrac{1}{2(n-1)}\sum\limits_{i=1}^{n-1} (x_{i+1}-x_{i})^{2}}{\dfrac{1}{n-1}\sum\limits_{i=1}^{n} (x_{i}-\bar{x})^{2}}=\frac{1}{2} \frac{\sum\limits_{i=1}^{n-1} (x_{i+1}-x_{i})^{2}}{\sum\limits_{i=1}^{n} (x_{i}-\bar{x})^{2}}\,,
	\end{equation}
	where $\bar{x}$ is the mean value of the magnitudes.
	If the sample size $n\geq 20$, $q$ is distributed approximately normally with a mean at 1.0, and with a variance of $\frac{n-2}{(n-1)(n+1)}$ \citep{24238,strunov2006applying}. The Allan variance alone was used for testing variability of extra-galactic sources \citep[e.g.,][]{2003A&A...403..105F,2018A&A...618A..80G,2018A&A...611A..52T}. Abb\'{e}'s criterion is for unevenly sampled data which in general are obtained by observations using ground based telescopes. This is a simple and effective method for analysing an unevenly sample of astronomical observations as described in \citet{2011A&A...536A..60S}, \citet{2013ARep...57..128M}. In Abb\'{e}'s test, the critical point is defined as 
	$q_{\rm c}=1+u_{\alpha}/\sqrt{n+0.5(1+u_{\alpha})^{2}}$, where $u_{\alpha}$ is 
	quantile of normal distribution for the significance level $\alpha$. The hypothesis about stochastic independence of the sample units is accepted under $q > q_{\rm c}$, otherwise the elements of the sample cannot be accepted as random and independent. In our case the sample consists of differences of magnitudes of comparison stars (A and B) and target blazars. Abb\'{e}'s statistics which correspond to that data are $q_{\rm A}$, and $q_{\rm B}$. If $q_{\rm A}$, and $q_{\rm B}$ are lower than $q_{\rm C}$, for the significance level $\alpha=0.001$, we conclude that there are statistically significant systematic variations present in the data.
	
	\subsection{F--test}\label{subsec3.2}
	We used F–test to determine the existence of brightness variability in the sample of blazars by following the method descibed in \citep[e.g.][and references therein] {2010AJ....139.1269D,2017MNRAS.465.4423G,2019SerAJ.199...55J}. We investigated variances of two data sets $X$, and $Y$ to test if they are equal to each other. The tested hypothesis is $H_{0}:Var X=Var Y$, and alternative $H:Var X>Var Y$. The corresponding statistic is
	
	\begin{equation}
	F = \frac{Var X}{Var Y}\,\cdot
	\end{equation}
	\noindent
	To implement the F-test to our sample, we calculate statistics: $F_{A}$, $F_{B}$, and its ratio $F_{A/B}$. The indices of statistics correspond to the data sets which are tested. $X$ refers to the differences of magnitudes of targets and comparison stars A, or B, the statistics are $F_{A}$, or $F_{B}$, respectively. $Y$ refers to the differences of magnitudes of comparison stars. The three $F_{A,\,B,\,A/B}$ statistics are compared with the critical values. The $F_{A/B}$ value should be $\sim$1, because it is expected that the tested brightness should be variable in the same manner for both comparison stars (A and B). When the $F_{A}$ and $F_{B}$ values are greater than critical ones (which correspond to the significance level 0.001, and number of freedom $n-1$, where $n$ is the sample size), the null hypothesis (of non variability) is discarded.
	
	\subsection{Amplitude of variability}
	\noindent
	The percentage of magnitude variations can be calculated by using the variability amplitude parameter ($VAP$), which was introduced by \citet{1996A&A...305...42H} and defined as
	\begin{equation}
	VAP = 100\sqrt{(M_{\rm MAX}-M_{\rm MIN})^{2}-2\sigma^{2}}\,(\%)\,,
	\end{equation}
	where $M_{\rm MAX}$ and $M_{\rm MIN}$ are the maximum, and minimum magnitude of the sources, and $\sigma$ is the average measurement error. \\
	\\
	The statistical results are listed in Table \ref{table:2}. The columns are: name of sources, band, number of data, results of Abb\'{e} criterion, results of F--test, maximum ($M_{\rm MAX}$), minimum ($M_{\rm MIN}$), average ($M_{\rm AV}$) magnitudes, amplitudes of full observations ($A=M_{\rm MAX}-M_{\rm MIN}$), and variability amplitude parameter ($VAP$).
	
	
	\begin{table*}
		\caption{Statistical results of objects variability.}  
		\label{table:2} 
		\centering 
		\resizebox{2.\columnwidth}{!}{
			\begin{tabular}{c c c c c c c c c c c}  
				\hline\hline   
				Name & Band & $n$ & Abb\'{e}'s criterion & F--test & $M_{\rm MAX}$ & $M_{\rm MIN}$ & $M_{\rm AV}\pm\sigma_{M}$ & $A$ & $VAP$ & Variable\\
				& & & $q_{A}$, $q_{B}$, $q_{c}$ & $F_{1}$, $F_{2}$, $F_{3}$, $F_{c}$ & (mag) & (mag) & (mag) & (mag) & \% &\\\hline
				0049+003 & $V$ & 30 & 0.18 , 0.15 , 0.48 & 1.30 , 20.64 , 15.92 , 3.29 & 16.731 & 16.166 & 16.461 $\pm$0.185 & 0.565 & 56.40 & V \\
				& $R$ & 40 & 0.15 , 0.15 , 0.54 & 1.23 , 48.16 , 39.14 , 2.76 & 16.292 & 15.835 & 16.100 $\pm$0.147 & 0.457 & 45.61 & V \\
				0907+336 & $V$ & 36 & 0.18 , 0.19 , 0.52 & 1.05 , 11.54 , 11.02 , 2.93 & 16.704 & 15.899 & 16.226 $\pm$0.180 & 0.805 & 80.51 & V \\
				& $R$ & 39 & 0.13 , 0.09 , 0.54 & 1.05 , 25.25 , 24.10 , 2.80 & 16.445 & 15.559 & 15.911 $\pm$0.191 & 0.886 & 88.56 & V \\
				1034+574 & $V$ & 47 & 0.20 , 0.21 , 0.57 & 1.00 , 83.77 , 83.96 , 2.54 & 16.919 & 15.545 & 16.086 $\pm$0.335 & 1.374 & 137.16 & V \\
				& $R$ & 47 & 0.23 , 0.22 , 0.57 & 1.01 , 96.54 , 97.13 , 2.54 & 16.504 & 15.253 & 15.744 $\pm$0.328 & 1.251 & 124.82 & V \\
				1212+467 & $V$ & 49 & 0.23 , 0.23 , 0.58 & 1.02 , 51.11 , 50.25 , 2.49 & 18.150 & 17.282 & 17.645 $\pm$0.203 & 0.868 & 86.13 & V \\
				& $R$ & 49 & 0.19 , 0.17 , 0.58 & 1.06 , 36.06 , 33.89 , 2.49 & 17.900 & 17.181 & 17.499 $\pm$0.186 & 0.719 & 71.54 & V \\
				1242+574 & $V$ & 43 & 0.25 , 0.26 , 0.56 & 1.04 , 28.77 , 27.66 , 2.66 & 18.167 & 17.371 & 17.710 $\pm$0.223 & 0.796 & 78.85 & V \\
				& $R$ & 51 & 0.24 , 0.27 , 0.59 & 1.10 , 58.77 , 64.44 , 2.44 & 17.816 & 16.990 & 17.353 $\pm$0.229 & 0.826 & 82.01 & V \\
				1429+249 & $V$ & 33 & 0.49 , 0.51 , 0.50 & 1.32 , 3.49 , 4.61 , 3.09 & 17.614 & 17.134 & 17.417 $\pm$0.107 & -- & -- & NV \\
				& $R$ & 37 & 0.51 , 0.50 , 0.52 & 1.10 , 2.56 , 2.82 , 2.89 & 17.343 & 17.076 & 17.197 $\pm$0.073 & -- & -- & NV \\
				1535+231 & $V$ & 43 & 0.30 , 0.31 , 0.56 & 1.11 , 31.34 , 34.81 , 2.66 & 19.036 & 18.133 & 18.472 $\pm$0.233 & 0.903 & 89.81 & V \\
				& $R$ & 44 & 0.15 , 0.18 , 0.56 & 1.12 , 16.41 , 18.34 , 2.63 & 18.610 & 17.797 & 18.193 $\pm$0.214 & 0.813 & 80.68 & V \\
				1556+335 & $V$ & 41 & 0.43 , 0.57 , 0.55 & 1.18 , 2.80 , 2.37 , 2.73 & 17.581 & 17.350 & 17.459 $\pm$0.064 & -- & -- & NV \\
				& $R$ & 50 & 0.77 , 0.63 , 0.58 & 1.44 , 1.23 , 1.77 , 2.46 & 17.080 & 16.886 & 16.988 $\pm$0.052 & -- & -- & NV \\
				1607+604 & $V$ & 42 & 0.26 , 0.27 , 0.55 & 1.12 , 23.81 , 21.33 , 2.69 & 17.677 & 17.152 & 17.400 $\pm$0.127 & 0.525 & 52.18 & V \\
				& $R$ & 48 & 0.38 , 0.41 , 0.58 & 1.23 , 8.55 , 6.97 , 2.51 & 17.140 & 16.747 & 16.956 $\pm$0.095 & 0.393 & 39.02 & V \\
				1612+378 & $V$ & 31 & 0.15 , 0.15 , 0.49 & 1.27 , 12.90 , 10.16 , 3.22 & 17.128 & 16.686 & 16.895 $\pm$0.137 & 0.442 & 44.16 & V \\
				& $R$ & 36 & 0.11 , 0.10 , 0.52 & 1.02 , 13.16 , 12.93 , 2.93 & 16.661 & 16.271 & 16.474 $\pm$0.111 & 0.390 & 38.94 & V \\
				1722+119 & $V$ & 36 & 0.11 , 0.12 , 0.52 & 1.05 , 202.59 , 192.64 , 2.93 & 16.780 & 14.888 & 15.571 $\pm$0.467 & 1.892 & 189.06 & V \\
				& $R$ & 40 & 0.11 , 0.11 , 0.54 & 1.00 , 1389.46 , 1387.88 , 2.76 & 16.343 & 14.371 & 15.083 $\pm$0.477 & 1.972 & 197.16 & V \\
				1741+597 & $V$ & 55 & 0.26 , 0.27 , 0.60 & 1.05 , 40.03 , 41.84 , 2.36 & 18.435 & 16.837 & 17.975 $\pm$0.313 & 1.598 & 159.37 & V \\
				& $R$ & 62 & 0.21 , 0.21 , 0.62 & 1.03 , 33.84 , 34.79 , 2.24 & 18.145 & 16.447 & 17.513 $\pm$0.310 & 1.698 & 169.71 & V \\
				\hline      
				\multicolumn{11}{l}{Notes. In the Variable column, V represents variable, and NV nonvariable source.}
			\end{tabular}
		}
	\end{table*}
	
	\subsection{Colour variability}
	In addition to the brightness variations, we also estimated colour variations of the blazars with respect to total duration of observations, and also with respect to $R$ magnitude. Studying colour variations is of importance to characterise the nature of the variations in the blazars which help explain the dominant emission mechanism. For all the blazars, we generate diagrams of colour $V-R$ with respect to $R$ magnitude (colour magnitude diagram (CMD)), and colour ($V-R$) with respect to time (Julian days). For linear regression, Pearson's correlation coefficient with null hypothesis probability were estimated on the data of these CMDs and colour verses time plots. The positive slope, of linear regression of the data in colour-magnitude diagram, is an indication of bluer when brighter (BWB), and negative are indications of redder when brighter (RWB) trend. The colour versus time and colour versus magnitude diagrams are presented for the most variable sources in Figs.~\ref{fig:VRT_exapmle}, and \ref{fig:VRR_exapmle}, and for the remaining sources in Appendices~\ref{app2}, \ref{app3}. The coefficients of linear regression (slope and intercept) and Pearson's correlation coefficients with probability are given in Tables \ref{table:3}, and \ref{table:4}, for colour-time and colour-magnitude diagrams, respectively. 
	If Pearson's coefficient $r$ is positive and probability (of no correlation) $P$ is lower than 0.05, we assume that the BWB colour-magnitude variation in source is present, if $r$ is negative (and $P<0.05$) we consider that RWB variation is present. In case of $P>0.95$, we can conclude that the correlation is not present in colour-magnitude data. In other cases we can not conclude anything about behaviour of the process. In the similar manner as in subsections \ref{subsec3.1} and \ref{subsec3.2} we tested colour $V-R$ indices of objects control stars, and the results are presented in paper \citet{2023PASRB..J} - in Press.
	
	\begin{figure}
		\centering
		\includegraphics[width=\columnwidth]{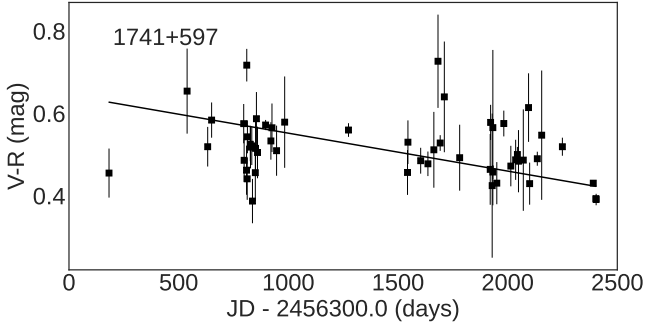}
		\caption{The light curve of colour indices $V-R$ variability during period July 2013 -- August 2019 of 1741+597. Details about the colour variability of all sources can be found in Table \ref{table:3}. }
		\label{fig:VRT_exapmle}
	\end{figure}
	%
	
	\begin{figure}
		\centering
		\includegraphics[width=\columnwidth]{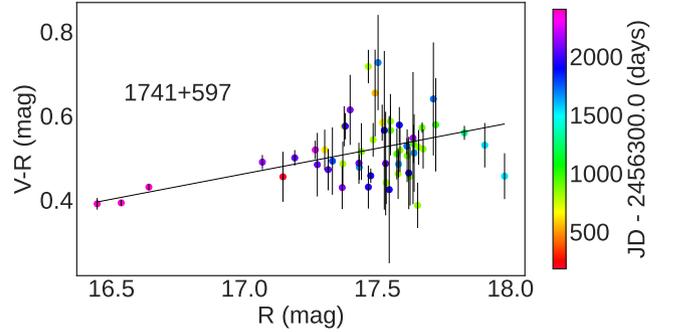}
		\caption{The correlation between colour indices $V-R$ and $R$-band magnitude of 1741+597. The colour bars indicate the progression of time. Details about the colour-magnitude correlations for all sources can be found in Table \ref{table:4}. }
		\label{fig:VRR_exapmle}
	\end{figure}

	\begin{table}
		\caption{The colour variations with respect to time.}  
		\label{table:3} 
		\centering     
		\resizebox{\columnwidth}{!}{%
			\begin{tabular}{c r c r c }  
				\hline\hline    
				Source & Slope & Intercept & r & P \\
				& ($\times10^{-5}$) & & & \\\hline   
				0049+003 & 4.0 $\pm$ 0.4 & 0.331 $\pm$ 0.007 & 0.52 & 3.10$\times10^{-3}$ \\
				0907+336 & -4.7 $\pm$ 0.5 & 0.373 $\pm$ 0.005 & -0.53 & 6.00$\times10^{-4}$ \\
				1034+574 & 1.0 $\pm$ 0.3 & 0.322 $\pm$ 0.004 & 0.16 & 2.82$\times10^{-1}$\\
				1212+467 & -2.3 $\pm$ 1.2 & 0.166 $\pm$ 0.012 & -0.15 & 2.97$\times10^{-1}$\\
				1242+574 & -0.3 $\pm$ 0.5 & 0.393 $\pm$ 0.010 & -0.03 & 8.20$\times10^{-1}$\\
				1429+249 & 0.5 $\pm$ 0.5 & 0.205 $\pm$ 0.009 & 0.06 & 7.19$\times10^{-1}$\\
				1535+231 & -2.1 $\pm$ 1.5 & 0.277 $\pm$ 0.027 & -0.12 & 4.54$\times10^{-1}$\\
				1556+335 & 0.4 $\pm$ 0.5 & 0.465 $\pm$ 0.009 & 0.06 & 7.34$\times10^{-1}$\\
				1607+604 & 4.1 $\pm$ 0.5 & 0.375 $\pm$ 0.007 & 0.58 & 1.00$\times10^{-4}$\\
				1612+378 & 1.3 $\pm$ 0.5 & 0.417 $\pm$ 0.006 & 0.23 & 1.72$\times10^{-1}$\\
				1722+119 & 0.1 $\pm$ 0.3 & 0.427 $\pm$ 0.005 & 0.02 & 8.77$\times10^{-1}$\\
				1741+597 & -9.2 $\pm$ 0.6 & 0.644 $\pm$ 0.012 & -0.79 & 1.46$\times10^{-12}$\\\hline 
				\multicolumn{5}{l}{Notes. Slope, and Intercept of $V-R$ against JD-2456300 days,} \\
				\multicolumn{5}{l}{r - Pearson's coefficient and P - null hypothesis probability.}\\
				
			\end{tabular}
		}
	\end{table}

	\begin{table}
		\caption{The colour – magnitude dependencies.}  
		\label{table:4} 
		\centering     
		\resizebox{\columnwidth}{!}{%
			\begin{tabular}{c r r r c }  
				\hline\hline    
				Source & Slope & Intercept & r & P \\\hline   
				0049+003 & 0.189 $\pm$ 0.018 & -2.64 $\pm$ 0.29 & 0.56 & 1.40$\times10^{-3}$ \\
				0907+336 & -0.093 $\pm$ 0.013 & 1.81 $\pm$ 0.20 & -0.39 & 1.38$\times10^{-2}$ \\
				1034+574 & 0.033 $\pm$ 0.007 & -0.19 $\pm$ 0.11 & 0.27 & 6.67$\times10^{-2}$ \\
				1212+467 & -0.055 $\pm$ 0.036 & 1.11 $\pm$ 0.63 & -0.12 & 4.07$\times10^{-1}$ \\
				1242+574 & -0.013 $\pm$ 0.022 & 0.61 $\pm$ 0.38 & -0.04 & 8.11$\times10^{-1}$ \\
				1429+249 & 0.226 $\pm$ 0.059 & -3.68 $\pm$ 1.02 & 0.24 & 1.34$\times10^{-2}$ \\
				1535+231 & -0.082 $\pm$ 0.055 & 1.73 $\pm$ 1.01 & -0.13 & 4.11$\times10^{-1}$ \\
				1556+335 & -0.096 $\pm$ 0.060 & 2.11 $\pm$ 1.01 & -0.12 & 4.64$\times10^{-1}$ \\
				1607+604 & 0.302 $\pm$ 0.038 & -4.69 $\pm$ 0.64 & 0.56 & 1.00$\times10^{-4}$ \\
				1612+378 & 0.175 $\pm$ 0.022 & -2.44 $\pm$ 0.36 & 0.61 & 1.00$\times10^{-4}$ \\
				1722+119 & 0.024 $\pm$ 0.005 & 0.08 $\pm$ 0.07 & 0.37 & 1.41$\times10^{-2}$ \\
				1741+597 & 0.121 $\pm$ 0.007 & -1.60 $\pm$ 0.12 & 0.86 & 1.43$\times10^{-16}$ \\\hline   
				\multicolumn{5}{l}{Notes. Slope, and Intercept of $V-R$ against $R$, r - Pearson's coefficient,}\\   
				\multicolumn{5}{l}{and P - null hypothesis probability.} \\
			\end{tabular}
		}
	\end{table}
	
	\subsection{Spectral variability}
	
	The flux density can be described by power law $F_{\nu}\varpropto\nu^{\alpha}$, where $\nu$ is frequency, and $\alpha$ is the spectral index. For the optical $V$, and $R$ bands, we calculated spectral index (similar to that presented for radio frequencies in paper \citet{2019A&A...630A..83Z}):
	
	\begin{equation}\label{eq:spec_ind_flux}
	\alpha = \frac{\log(F_{V}/F_{R})}{\log(\nu_{V}/\nu_{R})}\,,
	\end{equation}
	
	\noindent where $F_{V}$, and $F_{R}$ are fluxes of effective frequencies of $V$, and $R$ bands ($\nu_{V}$, and $\nu_{R}$), respectively. With flux magnitude relation equation (\ref{eq:spec_ind_flux}) can be written as:
	
	\begin{equation}
	\alpha = \frac{c-0.4(V-R)}{\log(\nu_{V}/\nu_{R})}\,,
	\end{equation}
	
	\noindent where $c=\log(ZP_{V}/ZP_{R})$, $ZP_{V}$, and $ZP_{R}$ are fluxes for magnitudes $V=0$, and $R=0$, respectively. The values $\nu_{V}$, $\nu_{R}$, $ZP_{V}$, and $ZP_{R}$ were taken from \citet{1998A&A...333..231B}.
	
	\noindent The uncertainty of the spectral index $\sigma_{\alpha}$ was calculated as in \citet{2019A&A...630A..83Z}.
	\begin{equation}
	\sigma_{\alpha} = \frac{1}{|\log(\nu_{V}/\nu_{R})|}\sqrt{(\sigma_{F_{V}}/F_{V})^{2}+(\sigma_{F_{R}}/F_{R})^{2}}\,,
	\end{equation}
	
	\noindent $\sigma_{F_{V}}$, and $\sigma_{F_{R}}$ are the uncertainties of flux densities at frequencies in $V$, and $R$ bands.
	\noindent Our sample consists of three FSRQs, one ISP, and four HSPs sources. Spectral indices $\alpha$ in the optical band against synchrotron peak frequency ($\log\nu$) are presented in Fig.~\ref{fig:SI_fr_example}, with triangles are marked FSRQs, circle ISP, and squares HSPs. Synchrotron peak frequencies of these sources were taken from \citet{2017ApJ...841..113M}, and \citet{2017A&A...598A..17C, 2019A&A...632A..77C}. \\
	\\
	The mean optical spectral index $\alpha$ is negative for FSRQ 1212+467 ($\alpha=-0.23\pm0.44$), lower than 1 for two HSPs 0907+336 ($\alpha=0.80\pm0.24$), and 1034+574 ($\alpha=0.82\pm0.20$), and greater than 1 for two FSRQs 1556+335 ($\alpha=1.67\pm0.30$), and 1612+378 ($\alpha=1.41\pm0.20$), ISP 1242+574 ($\alpha=1.18\pm0.45$), and two HSPs 1722+119 ($\alpha=1.40\pm0.15$), and 1741+597 ($\alpha=1.87\pm0.41$). For four sources (0049+003, 1429+249, 1535+231, and 1607+604) for which the synchrotron peak frequency is not available, their sub-class could not be determined, but the $\alpha$ is calculated and the average values are $1.22\pm0.30$, $0.16\pm0.36$, $0.35\pm0.73$, and $1.51\pm0.40$, respectively.
	\begin{figure}
		\centering
		\includegraphics[width=\columnwidth]{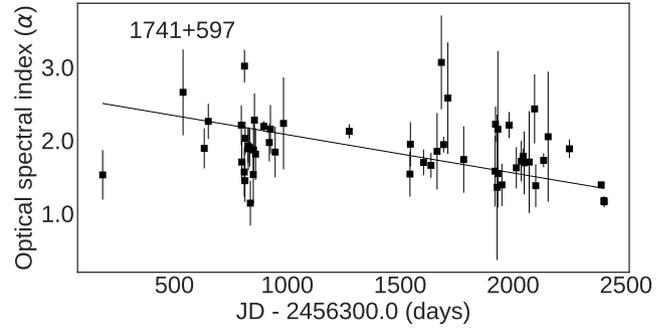}
		\caption{The light curve of optical spectral indices variability during period July 2013 -- August 2019 of 1741+597. Details about the $\alpha$ variability of all sources can be found in Table \ref{table:SIt}. }
		\label{fig:SIT_example}
	\end{figure}
	%
	
	\begin{figure}
		\centering
		\includegraphics[width=\columnwidth]{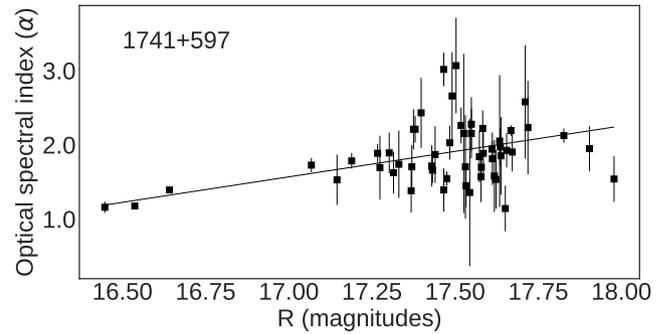}
		\caption{The correlation between optical spectral indices and $R$-band magnitude of 1741+597. Details about the $\alpha$-$R$ magnitude correlations for all sources can be found in Table \ref{table:SIR}. }
		\label{fig:SIR_example}
	\end{figure}
	%
	
	\begin{figure}
		\centering
		\includegraphics[width=\columnwidth]{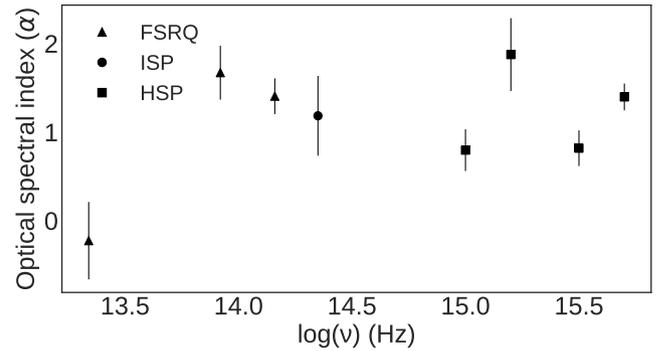}
		\caption{Average optical spectral index vs. synchrotron peak frequency. Sources 0049+003, 1429+249, 1535+231, and 1607+604 are excluded because their sub-class could not be determined. }
		\label{fig:SI_fr_example}
	\end{figure}

	\begin{table}
		\caption{The spectral index variations with respect to time.}  
		\label{table:SIt} 
		\centering     
		\resizebox{\columnwidth}{!}{%
			\begin{tabular}{c r c r c }  
				\hline\hline    
				Source & Slope & Intercept & r & P \\
				& ($\times10^{-5}$) & & & \\\hline   
				0049+003 & 2.3 $\pm$ 2.0 & 0.83 $\pm$ 0.04 & 0.52 & 3.30$\times10^{-3}$ \\
				0907+336 & -2.7 $\pm$ 3.0 & 1.06 $\pm$ 0.03 & -0.53 & 6.00$\times10^{-4}$ \\
				1034+574 & 5.0 $\pm$ 2.0 & 0.78 $\pm$ 0.03 & 0.15 & 3.06$\times10^{-1}$ \\
				1212+467 & -1.3 $\pm$ 7.0 & -0.11 $\pm$ 0.07 & -0.15 & 2.95$\times10^{-1}$ \\
				1242+574 & -2.0 $\pm$ 3.0 & 1.17 $\pm$ 0.06 & -0.03 & 8.31$\times10^{-1}$ \\
				1429+249 & 3.0 $\pm$ 3.0 & 0.11 $\pm$ 0.05 & 0.06 & 7.19$\times10^{-1}$\\
				1535+231 & -1.2 $\pm$ 9.0 & 0.52 $\pm$ 0.16 & -0.12 & 4.54$\times10^{-1}$\\
				1556+335 & 3.0 $\pm$ 3.0 & 1.59 $\pm$ 0.05 & 0.06 & 7.15$\times10^{-1}$\\
				1607+604 & 2.3 $\pm$ 3.0 & 1.08 $\pm$ 0.04 & 0.58 & 1.00$\times10^{-4}$\\
				1612+378 & 7.0 $\pm$ 2.0 & 1.31 $\pm$ 0.03 & 0.23 & 1.72$\times10^{-1}$\\
				1722+119 & 1.0 $\pm$ 2.0 & 1.37 $\pm$ 0.03 & 0.03 & 8.70$\times10^{-1}$\\
				1741+597 & -5.2 $\pm$ 3.0 & 2.60 $\pm$ 0.07 & -0.79 & 1.54$\times10^{-12}$\\\hline 
				\multicolumn{5}{l}{Notes. Slope, and Intercept of $\alpha$ against JD-2456300 days,}\\
				\multicolumn{5}{l}{r - Pearson's coefficient, and P - null hypothesis probability.}\\
				
			\end{tabular}
		}
	\end{table}

	\begin{table}
		\caption{The optical spectral index – magnitude dependencies.}  
		\label{table:SIR} 
		\centering     
		\resizebox{\columnwidth}{!}{%
			\begin{tabular}{c r r r c }  
				\hline\hline    
				Source & Slope & Intercept & r & P \\\hline   
				0049+003 & 1.07 $\pm$ 0.10 & -16.08 $\pm$ 1.64 & 0.56 & 1.40$\times10^{-3}$ \\
				0907+336 & -0.53 $\pm$ 0.07 & 9.21 $\pm$ 1.13 & -0.39 & 1.38$\times10^{-2}$ \\
				1034+574 & 0.19 $\pm$ 0.04 & -2.13 $\pm$ 0.61 & 0.27 & 6.68$\times10^{-2}$ \\
				1212+467 & -0.32 $\pm$ 0.20 & 5.32 $\pm$ 3.57 & -0.12 & 3.99$\times10^{-1}$ \\
				1242+574 & -0.07 $\pm$ 0.12 & 2.41 $\pm$ 2.17 & -0.04 & 8.08$\times10^{-1}$ \\
				1429+249 & 1.28 $\pm$ 0.34 & -21.93 $\pm$ 5.79 & 0.24 & 1.34$\times10^{-1}$ \\
				1535+231 & -0.46 $\pm$ 0.31 & 8.78 $\pm$ 5.71 & -0.13 & 4.11$\times10^{-3}$ \\
				1556+335 & -0.55 $\pm$ 0.34 & 10.97 $\pm$ 5.80 & -0.12 & 4.64$\times10^{-1}$ \\
				1607+604 & 1.71 $\pm$ 0.21 & -27.67 $\pm$ 3.62 & 0.56 & 1.00$\times10^{-4}$ \\
				1612+378 & 0.99 $\pm$ 0.13 & -14.91 $\pm$ 2.07 & 0.61 & 1.00$\times10^{-4}$ \\
				1722+119 & 0.13 $\pm$ 0.03 & -0.59 $\pm$ 0.43 & 0.36 & 1.72$\times10^{-2}$ \\
				1741+597 & 0.69 $\pm$ 0.04 & -10.10 $\pm$ 0.68 & 0.86 & 1.67$\times10^{-16}$ \\\hline  
				\multicolumn{5}{l}{Notes. Slope, and Intercept of $\alpha$ against $R$, r - Pearson's coefficient,}    \\
				\multicolumn{5}{l}{and P - null hypothesis probability.}    \\
			\end{tabular}
		}
	\end{table}

	\subsection{Spectral energy distribution (SED)}
	
	For nights when observations were obtained in the $B$, $V$, and $R$ bands, the calibrated magnitudes of 12 blazars were dereddened by subtracting Galactic extinction $A_{B,V,R}$ (see Table \ref{table:SED_gal}). The presented values of $A_{B,V,R}$ were calculated using the NASA Extragalactic Database Extinction calculator tool\footnote{\url{https://ned.ipac.caltech.edu/extinction_calculator}} (for $B$, $V$, and $R$ bands it is based on paper \citet{2011ApJ...737..103S}). The SEDs are compiled using extinction corrected flux densities ($F_{\nu}$) at $B$, $V$, and $R$ wavelengths. Fig. \ref{fig:SED_example} shows optical SED for 1741+597 for 9 different epochs. In Appendix \ref{app4} are presented optical SEDs for all sources, and the details (results of linear fits) are presented in Table \ref{table:SED}.\\

	\begin{table}
		\caption{The Galactic extinction.}  
		\label{table:SED_gal} 
		\centering     
		\begin{tabular}{c c c c }  
			\hline\hline    
			Source & $A_{B}$ (mag) & $A_{V}$ (mag) & $A_{R}$ (mag) \\\hline   
			0049+003&0.088&0.066&0.052\\
			0907+336&0.079&0.060&0.047\\
			1034+574&0.015&0.011&0.009\\
			1212+467&0.052&0.040&0.031\\
			1242+574&0.040&0.030&0.024\\
			1429+249&0.119&0.090&0.071\\
			1535+231&0.149&0.113&0.089\\
			1556+335&0.111&0.084&0.067\\
			1607+604&0.051&0.038&0.030\\
			1612+378&0.057&0.043&0.034\\
			1722+119&0.625&0.473&0.374\\
			1741+597&0.157&0.119&0.094\\\hline
			\multicolumn{4}{l}{Notes. $A_{B}$, $A_{V}$, and $A_{R}$ are galactic absorption for}  \\    
			\multicolumn{4}{l}{$B$, $V$, and $R$ bands.}  \\
		\end{tabular}
		
	\end{table} 
	
	\begin{figure}
		\centering
		\includegraphics[width=\columnwidth]{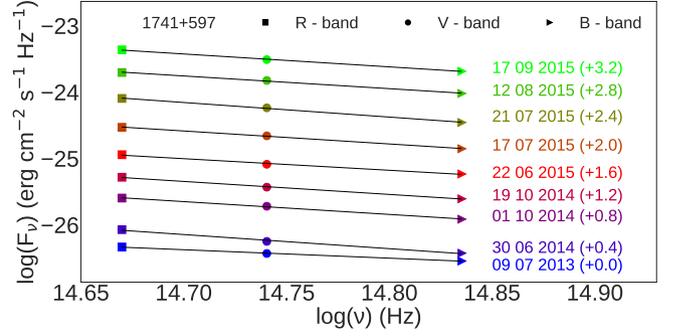}
		\caption{ The SED of 1741+597 in $B$, $V$, and $R$ bands. Details about the $\log F_{\nu}$-$\log \nu$ correlations for all sources can be found in Table \ref{table:SED}. }
		\label{fig:SED_example}
	\end{figure}

	\begin{table*}
		\caption{Straight-line Fits to Optical SEDs of 12 sources.}  
		\label{table:SED} 
		\centering 
		\resizebox{2.\columnwidth}{!}{
			\begin{tabular}{c c c c c c c c c c}  
				\hline\hline   
				\multicolumn{10}{c}{Source}\\
				Observation date & slope & intercept & r & P & Observation date & slope & intercept & r & P\\
				dd mm yyyy & & & & & dd mm yyyy & & & & \\\hline
				\multicolumn{10}{c}{0049+003}\\\hline
				06	09	2013	&	-1.081	$\pm$	0.077	&	-9.99	$\pm$	1.13	&	-0.998	&	0.045	&	15	08	2015	&	-1.530	$\pm$	0.169	&	-3.49	$\pm$	2.49	&	-0.994	&	0.070	\\
				19	12	2014	&	-1.260	$\pm$	0.175	&	-7.38	$\pm$	2.58	&	-0.990	&	0.088	&	13	09	2015	&	-1.665	$\pm$	0.306	&	-1.50	$\pm$	4.52	&	-0.983	&	0.116	\\\hline
				\multicolumn{10}{c}{0907+336}\\\hline
				14	04	2013	&	-0.862	$\pm$	0.003	&	-13.11	$\pm$	0.05	&	-1.000	&	0.003	&	22	05	2014	&	-0.982	$\pm$	0.061	&	-11.36	$\pm$	0.90	&	-0.998	&	0.039	\\
				01	03	2014	&	-1.033	$\pm$	0.016	&	-10.64	$\pm$	0.23	&	-1.000	&	0.010	&	21	10	2014	&	-1.016	$\pm$	0.101	&	-10.81	$\pm$	1.49	&	-0.995	&	0.063	\\\hline
				\multicolumn{10}{c}{1034+574}\\\hline
				09	07	2013	&	-1.129	$\pm$	0.075	&	-9.56	$\pm$	1.11	&	-0.998	&	0.043	&	22	05	2014	&	-1.162	$\pm$	0.073	&	-8.82	$\pm$	1.08	&	-0.998	&	0.040	\\
				01	03	2014	&	-1.255	$\pm$	0.040	&	-7.58	$\pm$	0.59	&	-0.999	&	0.020	&	19	02	2015	&	-1.282	$\pm$	0.188	&	-6.81	$\pm$	2.77	&	-0.989	&	0.093	\\\hline
				\multicolumn{10}{c}{1212+467}\\\hline
				09	07	2013	&	-0.097	$\pm$	0.133	&	-25.02	$\pm$	1.96	&	-0.591	&	0.598	&	24	12	2014	&	0.082	$\pm$	0.203	&	-27.75	$\pm$	3.00	&	0.374	&	0.756	\\
				01	04	2014	&	-0.182	$\pm$	0.106	&	-23.84	$\pm$	1.57	&	-0.864	&	0.336	&	21	02	2015	&	-0.112	$\pm$	0.121	&	-24.81	$\pm$	1.79	&	-0.678	&	0.526	\\
				27	06	2014	&	-0.112	$\pm$	0.161	&	-24.96	$\pm$	2.38	&	-0.571	&	0.613	&																\\\hline
				\multicolumn{10}{c}{1242+574}\\\hline
				02	04	2014	&	-1.160	$\pm$	0.055	&	-9.42	$\pm$	0.82	&	-0.999	&	0.030	&	04	07	2014	&	-1.392	$\pm$	0.157	&	-6.13	$\pm$	2.31	&	-0.994	&	0.071	\\
				22	05	2014	&	-1.151	$\pm$	0.131	&	-9.71	$\pm$	1.94	&	-0.994	&	0.072	&	25	12	2014	&	-1.274	$\pm$	0.033	&	-7.78	$\pm$	0.48	&	-1.000	&	0.016	\\
				28	06	2014	&	-1.359	$\pm$	0.087	&	-6.62	$\pm$	1.29	&	-0.998	&	0.041	&	14	05	2015	&	-1.756	$\pm$	0.221	&	-0.60	$\pm$	3.27	&	-0.992	&	0.080	\\\hline
				\multicolumn{10}{c}{1429+249}\\\hline
				04	04	2014	&	-2.657	$\pm$	2.310	&	12.96	$\pm$	34.08	&	-0.755	&	0.456	&	25	12	2014	&	-2.954	$\pm$	2.139	&	17.37	$\pm$	31.55	&	-0.810	&	0.399	\\
				28	06	2014	&	-2.853	$\pm$	2.328	&	15.85	$\pm$	34.33	&	-0.775	&	0.436	&	15	04	2015	&	-2.858	$\pm$	2.243	&	15.91	$\pm$	33.08	&	-0.787	&	0.424	\\
				04	07	2014	&	-2.789	$\pm$	2.390	&	14.89	$\pm$	35.24	&	-0.759	&	0.451	&	16	07	2015	&	-3.063	$\pm$	2.259	&	18.93	$\pm$	33.31	&	-0.805	&	0.405	\\\hline
				\multicolumn{10}{c}{1535+231}\\\hline
				04	04	2014	&	-0.725	$\pm$	0.118	&	-16.26	$\pm$	1.75	&	-0.987	&	0.103	&	25	12	2014	&	-0.757	$\pm$	0.047	&	-15.74	$\pm$	0.70	&	-0.998	&	0.040	\\
				25	05	2014	&	-0.413	$\pm$	0.172	&	-20.87	$\pm$	2.54	&	-0.923	&	0.251	&	12	07	2015	&	-0.733	$\pm$	0.109	&	-15.94	$\pm$	1.61	&	-0.989	&	0.094	\\
				27	06	2014	&	-0.654	$\pm$	0.277	&	-17.34	$\pm$	4.08	&	-0.921	&	0.255	&	18	07	2015	&	-0.799	$\pm$	0.235	&	-15.01	$\pm$	3.47	&	-0.959	&	0.182	\\\hline
				\multicolumn{10}{c}{1556+335}\\\hline
				04	04	2014	&	-1.441	$\pm$	0.046	&	-5.10	$\pm$	0.68	&	-0.999	&	0.020	&	20	04	2015	&	-1.483	$\pm$	0.063	&	-4.51	$\pm$	0.94	&	-0.999	&	0.027	\\
				27	06	2014	&	-1.354	$\pm$	0.084	&	-6.38	$\pm$	1.24	&	-0.998	&	0.039	&	12	07	2015	&	-1.613	$\pm$	0.069	&	-2.61	$\pm$	1.02	&	-0.999	&	0.027	\\
				04	07	2014	&	-1.514	$\pm$	0.017	&	-4.05	$\pm$	0.25	&	-1.000	&	0.007	&																\\\hline
				\multicolumn{10}{c}{1607+604}\\\hline
				08	07	2013	&	-1.025	$\pm$	0.044	&	-11.24	$\pm$	0.65	&	-0.999	&	0.028	&	03	07	2014	&	-0.967	$\pm$	0.010	&	-12.04	$\pm$	0.15	&	-1.000	&	0.007	\\
				01	03	2014	&	-0.729	$\pm$	0.100	&	-15.57	$\pm$	1.48	&	-0.991	&	0.087	&	18	10	2014	&	-1.203	$\pm$	0.087	&	-8.62	$\pm$	1.29	&	-0.997	&	0.046	\\
				28	05	2014	&	-0.753	$\pm$	0.135	&	-15.19	$\pm$	1.99	&	-0.984	&	0.113	&	12	06	2015	&	-1.659	$\pm$	0.043	&	-1.96	$\pm$	0.63	&	-1.000	&	0.016	\\
				28	06	2014	&	-0.697	$\pm$	0.100	&	-16.01	$\pm$	1.47	&	-0.990	&	0.090	&	17	07	2015	&	-1.936	$\pm$	0.157	&	2.10	$\pm$	2.31	&	-0.997	&	0.051	\\\hline
				\multicolumn{10}{c}{1612+378}\\\hline
				08	07	2013	&	-0.890	$\pm$	0.044	&	-12.98	$\pm$	0.64	&	-0.999	&	0.031	&	01	10	2014	&	-0.995	$\pm$	0.077	&	-11.44	$\pm$	1.13	&	-0.997	&	0.049	\\
				28	05	2014	&	-0.786	$\pm$	0.172	&	-14.50	$\pm$	2.54	&	-0.977	&	0.137	&	14	06	2015	&	-1.093	$\pm$	0.138	&	-10.04	$\pm$	2.03	&	-0.992	&	0.080	\\
				29	06	2014	&	-0.858	$\pm$	0.224	&	-13.44	$\pm$	3.30	&	-0.968	&	0.162	&	18	07	2015	&	-1.119	$\pm$	0.125	&	-9.66	$\pm$	1.85	&	-0.994	&	0.071	\\\hline
				\multicolumn{10}{c}{1722+119}\\\hline
				09	07	2013	&	-1.042	$\pm$	0.020	&	-10.01	$\pm$	0.29	&	-1.000	&	0.012	&	13	07	2015	&	-1.242	$\pm$	0.225	&	-7.13	$\pm$	3.32	&	-0.984	&	0.114	\\
				29	06	2014	&	-1.052	$\pm$	0.026	&	-9.75	$\pm$	0.39	&	-1.000	&	0.016	&	11	08	2015	&	-1.289	$\pm$	0.146	&	-6.49	$\pm$	2.15	&	-0.994	&	0.072	\\
				22	04	2015	&	-1.225	$\pm$	0.153	&	-7.52	$\pm$	2.26	&	-0.992	&	0.079	&	17	09	2015	&	-1.306	$\pm$	0.120	&	-6.27	$\pm$	1.77	&	-0.996	&	0.058	\\
				13	05	2015	&	-1.005	$\pm$	0.136	&	-10.75	$\pm$	2.00	&	-0.991	&	0.085	&																\\\hline
				\multicolumn{10}{c}{1741+597}\\\hline
				09	07	2013	&	-1.273	$\pm$	0.053	&	-7.67	$\pm$	0.78	&	-0.999	&	0.027	&	17	07	2015	&	-1.957	$\pm$	0.060	&	2.18	$\pm$	0.89	&	-1.000	&	0.020	\\
				30	06	2014	&	-2.129	$\pm$	0.181	&	4.75	$\pm$	2.68	&	-0.996	&	0.054	&	21	07	2015	&	-2.197	$\pm$	0.087	&	5.74	$\pm$	1.29	&	-0.999	&	0.025	\\
				01	10	2014	&	-1.931	$\pm$	0.085	&	1.93	$\pm$	1.25	&	-0.999	&	0.028	&	12	08	2015	&	-1.914	$\pm$	0.101	&	1.57	$\pm$	1.49	&	-0.999	&	0.034	\\
				19	10	2014	&	-1.958	$\pm$	0.075	&	2.24	$\pm$	1.11	&	-0.999	&	0.024	&	17	09	2015	&	-1.944	$\pm$	0.069	&	1.95	$\pm$	1.01	&	-0.999	&	0.022	\\
				22	06	2015	&	-1.747	$\pm$	0.142	&	-0.93	$\pm$	2.10	&	-0.997	&	0.052	&																\\\hline
				\multicolumn{10}{l}{Notes. Slope, and Intercept of $\log F_{\nu}$ against $\log \nu$, r - Pearson's coefficient and P - null hypothesis probability.}\\
				
			\end{tabular}
		}
	\end{table*}
	
	\section{Results of individual targets} \label{sec:4}
	Abb\'{e}'s criterion and F-test show that the objects are variable in $V$, and $R$ bands in relation to both comparison stars $\rm A$, and $\rm B$ with exception of two objects. The 1424+249 is considered to be variable in relation to a comparison star $\rm A$ according to Abb\'{e}'s criterion in both bands, and in relation to B in according to Abb\'{e}'s criterion in $R$ band, and F--test in both bands. The object 1556+335 is variable only in $V$ band in relation to star A. Magnitudes of sources 1212+467, 1242+574, 1535+231, and 1741+597 are not homogeneous in relation to the standard deviation. When the sources were fainter, the standard deviation was greater, and vice versa. Also, the colour of sources was tested using the same statistical tests, both tests did not show that the colour is variable. Optical variabilities of sources 0907+336, 1034+574, 1212+467, 1242+574, 1607+604, 1612+378, and 1722+119 in $B$, and $R$ bands were investigated by Abrahamyan et al. (2019). Their optical variability was classified as low. \\
	\\
	The data from 2013 to 2015 were part of the data which were used for analysing variability of the sources, and the results were presented in paper \citet{2018A&A...611A..52T}.
	For sources 1535+231, 1556+335, 1607+604, 1722+119, and 1741+597 data from 2016 to 2019 were used for testing comparison stars for differential photometry in \citet{2018PASRB..18...197J}, and for obtaining their long-term period variability with Least Squares Method (LSM) iteratively in paper \citet{2019SerAJ.199...55J}, analogously   periodicity analysis for these blazars in short and long timescales was performed in \citet{2020BlgAJ..33...38J} as well as colour variability one in \citet{2020PASRB..20...23J}. For the same sources the data from 2013-2019 were used for obtaining the periods of short and long term variations with LSM iteratively which was presented in paper \citet{2020jsrs.conf...21D}. Moreover, the data (2013-2019) were used for testing the control stars for differential photometry (\citet{2021POBeo.100..253J}, and \citet{2023PASRB..J} - in Press). 
	
	\subsection{0049+003} 
	The source was first detected by HEAO 2 onboard of the EINSTEIN satellite \citep{1996yCat.9013....0H}. The large bright quasar survey identified it as a quasar through its spectrum and the redshift was found to be z = 0.399 \citep{1995AJ....109.1498H}. Later using another spectral analysis its redshift estimation was confirmed and found that z = 0.399714 \citep{2015ApJS..219...39R}. \citet{2007ApJS..171...61H} classified it as an FSRQ. The absolute magnitude of the source was estimated to be M$_{I} = -$25.48 \citep{2011A&A...525A..37M}. In paper \citet{2013ApJ...779..104J} source was catalogued as the hot dust-poor quasar with logarithm of the mass of the central black hole and the ratio of bolometric luminosity to Eddington luminosity 8.43 $\pm$ 0.01 M$_{\sun}$ and $0.959\pm0.030$, respectively, something similar was derived in paper \citet{2020ApJS..249...17R} 8.425803 $\pm$ 0.018190, and logarithmic Eddington ratio -0.183588. 
	In optical radio correlation study with optical data from SDSS and radio data from FIRST surveys, the optical/radio morphology of the object was classified as the optical/radio emission from the core of the source and extended radio jet emission in papers \citet{2004AJ....127.2565D}, and \citet{2011AJ....141..182K}. Comparing two epochs of FIRST survey with the higher angular resolution data of 1.4 GHz survey of SDSS Stripe 82, two diffuse lobes were visible on either side of the core
	and morphological class of source was defined as: core-lobe morphology (core is surrounded by two distinct non variable lobe components \citep{2013ApJ...769..125H}). \citet{2011A&A...528A..95G} investigate the optical variability in $r$ band using SDSS DR7 which released multi-epoch data covering about nine years. The source shows variation of $\Delta r =$ 0.44 mag. \\
	\\
	During our monitoring, the brightness changed by $\sim$0.5 magnitudes in $V$ and $R$ bands.	The colour of the blazar has changed by $\sim$0.2 magnitude during observational duration and shows BWB variations. The BWB variations can be seen in Fig. \ref{fig:FigColourMag} (colour-magnitude diagram) in Appendix~\ref{app3}. 
	
	\subsection{0907+336} 
	The source is also known as Ton 1015 and was for the first time noticed at the Tonantzintla Observatory in the second survey of blue stars in the north galactic pole, and its photographic magnitude was estimated to be 16$\pm$0.5 \citep{1959BOTT....2r...3C}. The source was detected in the radio band in a survey of faint sources at 5 GHz radio band by National Radio Astronomy Observatory (NRAO) \citep{1971AJ.....76..980D}. In the cross-identification of optical and radio sources, the object was classified as a BL Lac and the redshift was estimated z = 0.354 from spectrum \citep{2000ApJS..129..547B}. Its synchrotron peak frequency $\nu_{peak} = \rm{10}^{14.48}$ Hz was estimated and classified as an ISP, and radio to optical spectral index was found to be 0.28 in \citet{2016ApJS..226...20F}. In other studies the source is classified as an HSP \citep[e.g.][]{2006A&A...445..441N,2011ApJ...743..171A,2017ApJ...841..113M,2017A&A...598A..17C}. We classify the source as HSP according to the value for $\nu_{peak} = \rm{10}^{15.0}$ from \citet{2017A&A...598A..17C}. Using broad band SED modeling with synchrotron self-Compton (SSC)/Thomson model, its jet parameters were estimated \citep{2018ApJS..235...39C}. \\
	\\
	We noticed that in both bands the brightness decreases by $\sim$0.8 magnitude. A few outbursts in both bands occurred, three between 1 March 2014 and 16 May 2016, and one between 18 October 2017 and 4 October 2018. The colour also decreased by $\sim$0.2 magnitude during our observations. From colour-magnitude dependencies, we conclude that RWB variation is present, the colour index is smaller when the brightness of the blazar decreases, see Fig \ref{fig:FigColourMag} and Table \ref{table:4}. 
	
	\subsection{1034+574} The source was discovered during Green Bank 4.85 GHz survey with NRAO 91 m telescope. The telescope was used for three surveys in 1986, 1987, and 1988, and two catalogues which contain this object were published in \citet{1991ApJS...75....1B}, and \citet{1991ApJS...75.1011G}. The first time source was classified as BL Lac in paper \citet{1996A&A...309..419N}. The spectroscopic redshift $z=1.0957$, together with absolute $i$ magnitude -28.8, and the mass of the central black hole $10^{9.89655}$ $M_{\sun}$ were determined during the Large Sky Area Multi-Object Fibre Spectroscopic Telescope (LAMOST) Quasar Survey \citep{2018AJ....155..189D}. The classification of the source by synchrotron peak frequency position was discussed in a few papers. In the beginning the source was classified as ISP (\citet{2006A&A...445..441N}, and \citet{2011ApJ...743..171A})
	and later as HSP \citep{2016ApJS..226...20F,2017ApJ...841..113M,2019A&A...632A..77C}. We adopted for $log \nu_{peak} = 15.5$ value from \citet{2019A&A...632A..77C}, the 3HSP catalogue of extreme and high synchrotron peaked blazars. Physical parameters of the jet were estimated by \citet{2018ApJS..235...39C} using synchrotron self-Compton (SSC)/Thomson model. \\
	\\
	During imaging of host galaxies in $R$ band the source remains unresolved, only historical $R=15.99\pm0.03$ magnitude of the source core was recorded on 16 December 1998, by \citet{2003A&A...400...95N}. This object is one of the three, from our monitoring, with the highest brightness changes of about 1.3 magnitude. The colour has tendencies to change during time of observations (about 0.3 mag). From colour magnitude dependencies we can conclude that small BWB variations are present, which is one of the characteristics of BL Lac objects. During TJO monitoring one outburst was detected.
	
	\subsection{1212+467} 
	The source was discovered in 1400 MHz Green Bank radio sky survey \citep{1972AcA....22..227M}. In Roma-BZCAT Multifrequency Catalogue of Blazars, it was classified as a FSRQ \citep{2015Ap&SS.357...75M}. Its spectroscopic redshift was determined to be z = 0.720154 \citep{2015ApJS..219...39R}. Radio morphology of the source was found to be lobe-core-lobe \citep{2011AJ....141..182K}. In the catalogue of Spectral Properties of Quasars from SDSS DR14 \citep{2020ApJS..249...17R} are available the logarithmic fiducial single-epoch black hole mass calculated based on H$\beta$, Mg II and C IV lines (8.891813 $\pm$ 0.057461), and logarithmic Eddington ratio based on fiducial single-epoch black hole mass (-0.707690). The logarithm of $\nu_{peak}$ is estimated as 13.34 in \citet{2017ApJ...841..113M}. In paper \citep{2015AJ....149..203K} is given the spectral index difference (0.2) for the reddening law from \citet{2014ApJ...788..123L}. \\
	\\
	The different values of $V$, and $R$ magnitudes were given in several catalogues. The $R=17.13$ magnitude from catalogue of the CLASS blazar survey given by \citet{2001MNRAS.326.1455M} is close to the minimum value which we observed. With designation LQAC 183+046 007 source participates in the 1st and the 2nd Large Quasar Astrometric Catalogue which is a compilation of all the recorded quasars \citep{2009A&A...494..799S,2012A&A...537A..99S}. From the 1st LQAC catalogue $V=17.77$, and $R=17.42$ magnitudes are close to our average magnitudes. In the 2nd LQAC only $V=19.14$ magnitude was presented and this is the highest magnitude ever observed. In both bands the brightness changes by about 0.8 magnitude, from 2013 to 2019. The slope and Pearson's coefficient of colour--time and colour--magnitude dependencies are almost 0 with probability greater than 0.95. Colour values are in range of about 0.3 mag, around averaged value. We can not say that even tendencies are present because the slope and Pearson's coefficient are close to 0, but with probability less than 0.95, we can say that this object shows nearly achromatic behaviour.
	
	\subsection{1242+574} 
	The source was catalogued for the first time in the 87GB catalogue \citep{1991ApJS...75.1011G}. In the 12th edition of a catalogue of quasars and active nuclei, it was classified as BL Lac \citep{2006A&A...455..773V}. Its spectroscopy redshift z = 0.99822855 was estimated \citep{2015ApJS..219...39R}. The source $\nu_{peak}$ = 14.35 Hz in the observed frame ($\nu f_{\nu}$), so is an ISP blazar \citep[e.g.,]{2011ApJ...743..171A,2017ApJ...841..113M}. The source is in 1st and 3rd Fermi-LAT catalogues of sources above 10 GeV \citep{2013ApJS..209...34A,2015ApJS..218...23A}. In the MST catalogue of $\gamma$-ray source candidates above 10 GeV the source has designation 9Y-MST J1244+5709 \citep{2018A&A...619A..23C}. \\
	\\
	In both bands the brightness change is 0.8 magnitude. Similarly as object 1212+467, we can say that this object shows nearly achromatic behaviour. The slope and Pearson's coefficient of colour--time and colour--magnitude dependencies are negative, close to 0 and probability is greater than 0.05 and less than 0.95. The colour has tendencies to change during time of observations (almost 0.4 mag).
	
	\subsection{1429+249} 
	The source was discovered in second MIT--Green Bank 5 GHz radio survey \citep{1990ApJS...72..621L}. With broad Balmer and other permitted lines in spectra, it was classified as a Seyfert 1 (SY 1) galaxy in paper \citep{2006A&A...455..773V}, and with general spectroscopic characteristics in paper \citep{2022ApJS..260...33S}. In an all sky catalog of $\gamma$-ray blazars, the source was classified as a dual nature of both BL Lac as well as FSRQ \citep{2014ApJS..215...14D}. Its spectroscopic redshift was determined to be z = 0.40659 \citep{2018ApJ...866...33L}. In the paper \citet{2020ApJS..249...17R} were provided the logarithmic black hole mass (8.658600 $\pm$ 0.027332), and logarithmic Eddington ratio (-0.853556), both calculated based on H$\beta$, Mg II and C IV lines. Absolute i band magnitude is -24.134 from \citep{2013ApJ...768...37C}. The spectral index difference of 0.006 is given in paper \citep{2015AJ....149..203K}.\\
	\\
	The source is known also as LQAC 217+024 010. In the 1st LQAC catalogue were given $V=16.09$, and $R=17.43$ magnitudes, and in the 2nd LQAC $V=17.68$, and $R=17.44$ \citep{2009A&A...494..799S,2012A&A...537A..99S}. The $V$ from the 1st catalogue is lower than in $R$ band (authors), and the remained values are out of range of our observed magnitudes. The brightness of the source changed by 0.5 and 0.3 magnitude during six years in $V$ and $R$ band, respectively. Abb\'{e}'s and F statistics for this object are close to the critical values. Abb\'{e}'s criterion shows that the object has systematic variations in relations to comparison star $A$ in $V$ band, and to both stars in $R$ band. F-test shows that the object is variable only in $V$ band. From the colour--magnitude relations we can say that the BWB variations are present.
	
	\subsection{1535+231} The authors of papers \citet{2001ApJ...549..780A}, and \citet{2001ApJ...553L..11A} claim that the object is correlated to the nearby active galaxy Arp 220\footnote{ Also known as IC 4553 - galaxy merging system with two nuclei.} ($z=0.018$) and most likely has been ejected from it, even the object is at 43.1 arcmin distance from the galaxy, and has higher redshift ($z=0.4627$). Again in 2015 the redshift was determined by spectroscopy $z=0.462515$, in \citet{2015ApJS..219...39R}, when the object was classified as QSO. Based on mid-infrared colours of Wide-Field Infrared Survey Explorer the source was classified as mixed BL Lac and FSRQ blazar \citep{2019ApJS..242....4D}. The source was classified according to its general spectroscopic characteristics as Sy1 in paper \citet{2022ApJS..260...33S}. The logarithmic black hole mass (8.399292 $\pm$ 0.047624), and logarithmic Eddington ratio (-0.932017), both were calculated based on H$\beta$, Mg II and C IV lines, were provided in the paper \citet{2020ApJS..249...17R}. The spectral index difference according to \citet{2015AJ....149..203K} is 0.024. \\
	\\
	In both bands the brightness changed by about 0.9 magnitudes.
	The colour changed during time of observations almost a half magnitude. In case of colour--magnitude relations we can not say that RWB variations are present, only the RWB tendencies are present because the probability is greater than 0.05. This object is the faintest (average magnitudes are greater than 18 mag in both filters), which differs from the historical $V=17.7$ given in \citet{2001A&A...374...92V}.
	
	\subsection{1556+335} The source was detected for the first time during NRAO 5 GHz radio survey of faint sources, which was initiated in 1967 and presented in \citet{1971AJ.....76..980D}. It was identified as QSO by \citet{1979ApJS...41..689W}, later as FSRQ by \citet{2015Ap&SS.357...75M} in the 5th edition of the Roma-BZCAT Multifrequency Catalogue of Blazars. The first spectroscopic redshift $z=1.65$ by \citet{1979ApJS...41..689W} is similar to the one later determined by \citet{2015ApJS..219...39R}, $z=1.653598$. The presence of two absorption complexes in the spectrum could be explained with one of two models: one in which the source is directly responsible for velocities seen in both complexes, and the second in which complexes are related with two clusters, one cluster contains the source, while the other one is in the line of sight \citep{1986ApJ...310...40M}. The source radio morphology class was defined as \textit{core} - a quasar with radio emission only at the optical position, in \citet{2011AJ....141..182K}. Using spectral properties of quasar the logarithmic black hole mass (10.024996 $\pm$ 0.046142), and logarithmic Eddington ratio (-0.874986), were provided in the paper \citet{2020ApJS..249...17R}. The $\nu_{peak}$ is 13.92 \citep{2017ApJ...841..113M}.\\
	\\
	During our monitoring the object is variable only in $V$ band in relation to star A. This is the most stable object from our list. The historical $V=17$ magnitude, the lowest ever detected, was given in catalogue \citet{1987ApJS...63....1H}, and $R=16.94$ (for 1996.523) in paper \citet{2001AJ....121.1872H} is close to the average magnitude which we observed. For six years the brightness decreased by 0.2 mag in both bands. From the colour--time and colour--magnitude dependencies we can conclude that colour had not changed during time and that the achromatic behaviour is present. 
	
	\subsection{1607+604} After NRAO 4.85 GHz survey the source was catalogued by \citet{1991ApJS...75.1011G}, and \citet{1991ApJS...75....1B}, in second paper source was marked as extended. The redshift and classification as radio-loud quasar are obtained by spectroscopy in \citet{1998ApJS..118..127L}. The authors of \citet{2014ApJS..215...14D} classified the object as BL Lac. The radio and optical cross-identification of the source was accomplished by authors of paper \citet{2000ApJS..129..547B}. They presented the source redshift $z=0.178$, and radio emission as extended and resolved into three components. \\
	\\
	During time of observations the brightness changes by 0.5 and 0.4 magnitudes in $V$ and $R$ bands, respectively. The colour has tendencies to change by about 0.4 mag. In case of colour--magnitude relations we can say that BWB variations are present.
	
	\subsection{1612+378} In the 5th edition of the Roma-BZCAT Multifrequency Catalogue of Blazars the source was classified as FSRQ. The redshift $z=1.531239$ determined by spectroscopy was given in \citet{2015ApJS..219...39R}. Absolute $i$ magnitude of -28.332 mag is obtained in paper \citet{2011ApJS..194...42R}. As 1556+335, the radio morphology was classified as \textit{core} in \citet{2011AJ....141..182K} and the logarithmic black hole mass (9.684895 $\pm$ 0.084033), and logarithmic Eddington ratio (-0.454582), were provided in the paper \citet{2020ApJS..249...17R}. The source is ISP, its synchrotron peak frequency is $log \nu_{peak} = 14.16$, derived in \citet{2017ApJ...841..113M}. \\
	\\
	In both bands amplitudes of the brightness changes are 0.4 mag. The amplitude of colour changes is about 0.2 mag. In case of colour--magnitude relations we can say that BWB variations are present, which is one of the characteristics of BL Lac objects.
	
	\subsection{1722+119} 
\begin{figure}
		\centering
		\includegraphics[width=\columnwidth]{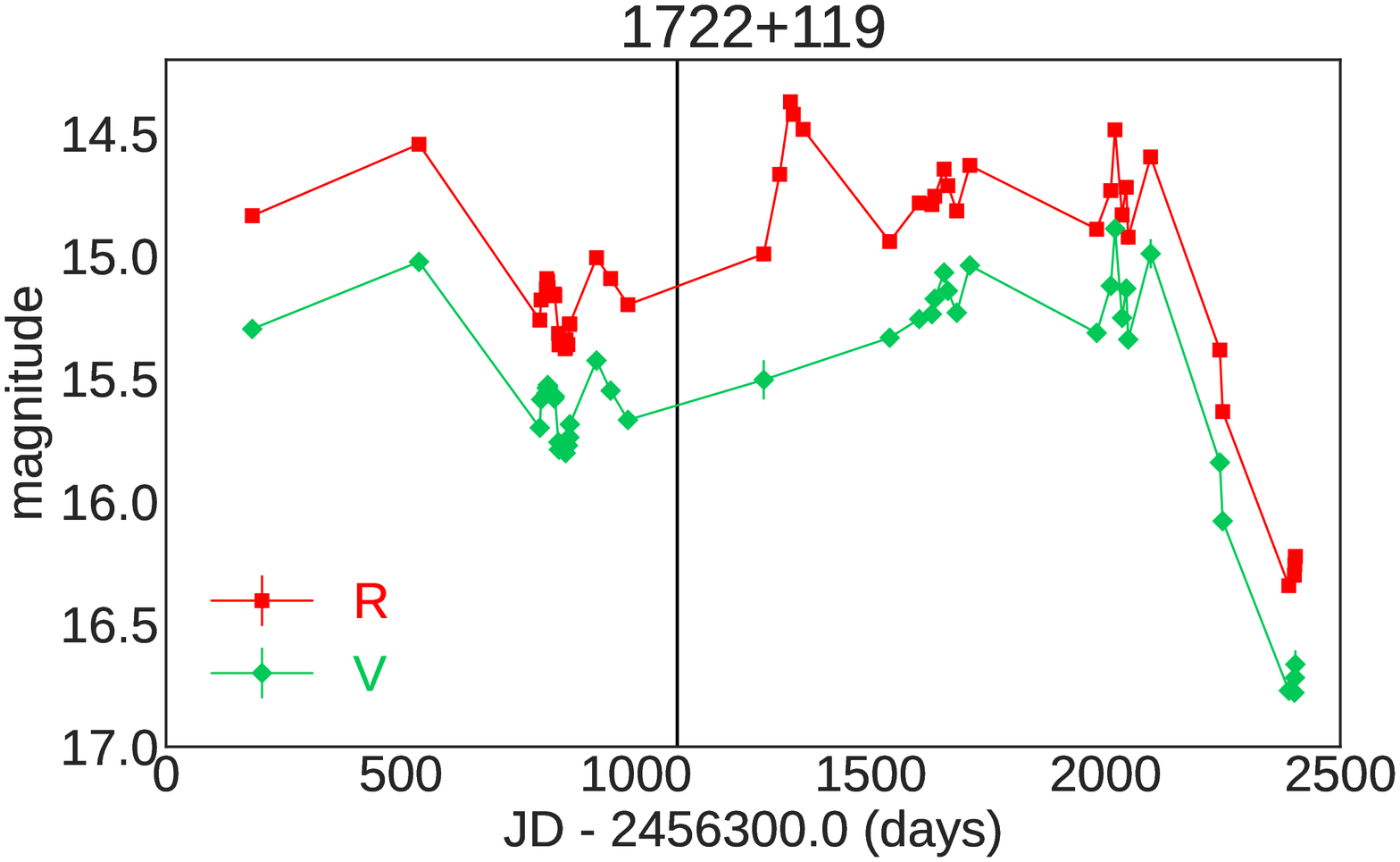}\\
		\includegraphics[width=0.48\hsize]{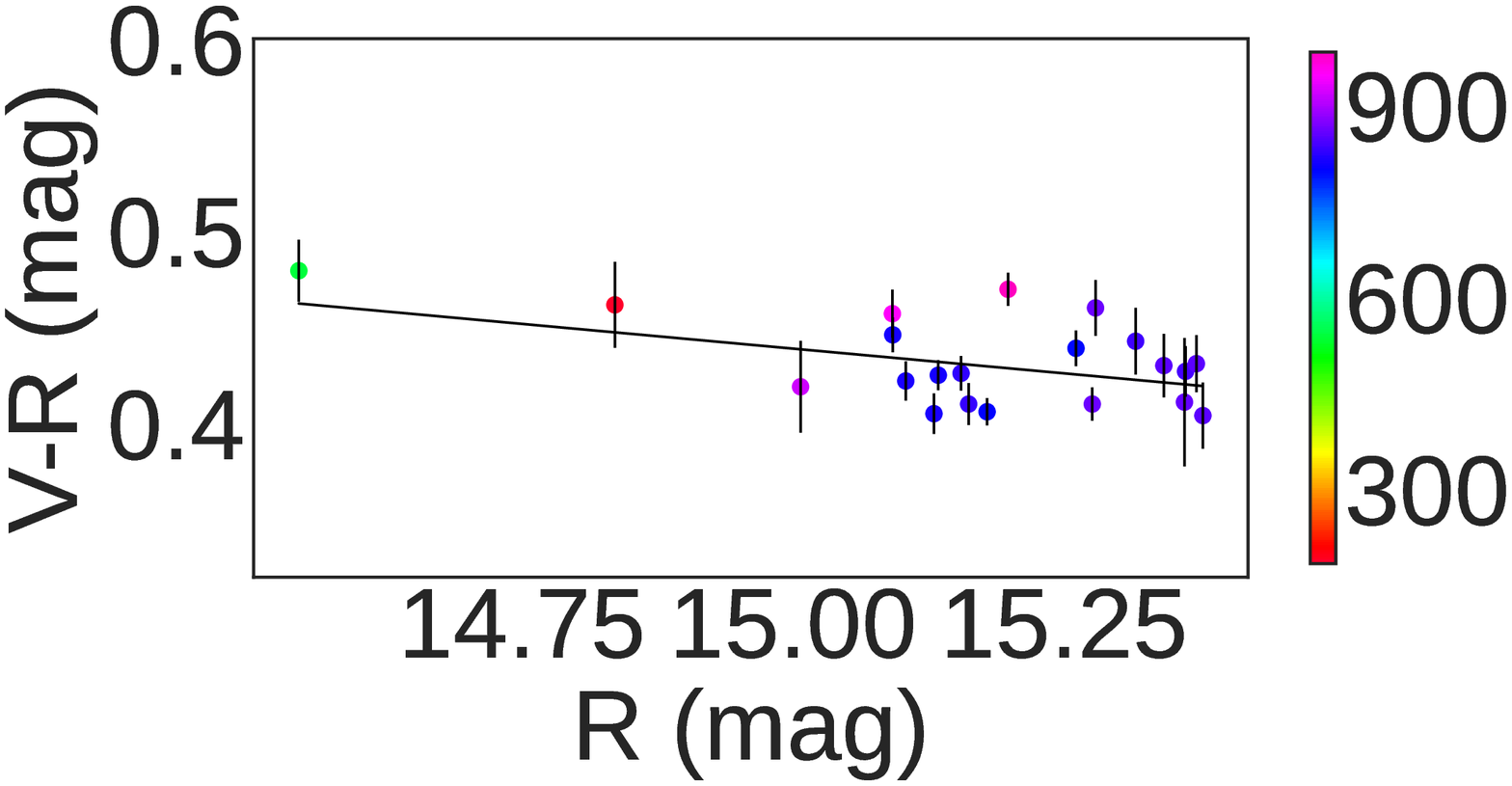}
		\includegraphics[width=0.48\hsize]{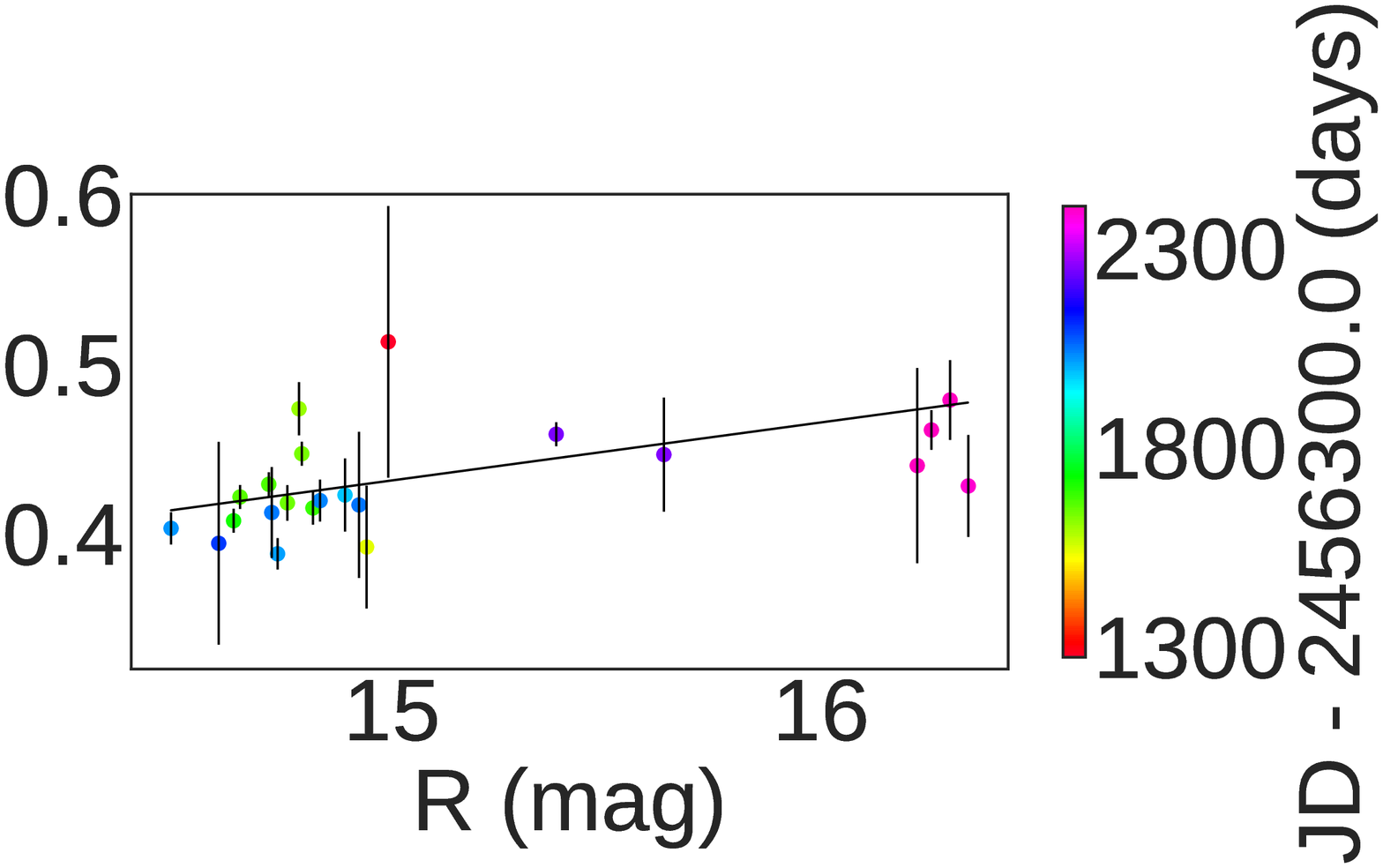}
		\caption{Light curve of 1722+119; colour-magnitude diagrams 2013--2016 (left-bottom), and 2016--2019 (right-bottom).
		}
		\label{fig:Fig1722_123}
	\end{figure}

This is one of the first discovered BL Lac objects. For the first time it appeared in the fourth Uhuru catalogue of X-ray sources \citet{1978ApJS...38..357F}. After a decade the object was independently classified as BL Lac, and its redshift estimation was given in papers \citet{1989MNRAS.240...33G} $z=0.018$, and \citet{1990ApJ...350..578B} $z>0.1$, and historical magnitude was $V=16.6$ mag on 1979, \citet{1989MNRAS.240...33G}.	In \citet{2016MNRAS.459.3271A} new redshift was given $0.34\pm 0.15$. The object is the one of the sources detected with MAGIC in TeV, announced by \citet{2013ATel.5080....1C}. The MAGIC observations were triggered by the optical outburst on May 2013 when the $R$ band magnitude reached 14.65 mag which was the biggest ever observed since 2005, when the Tuorla blazar monitoring programme started. According to the position of the synchrotron peak frequency object is HSP confirmed by \citet{2006A&A...445..441N}, \citet{2011ApJ...743..171A}, and \citet{2019A&A...632A..77C}. \citet{2019A&A...632A..77C} included this source in the third catalogue of extremely and high synchrotron peaked blazars, with $log \nu_{peak} = 15.7$. Jet properties of source were analysed but only the core temperature was obtained to be higher than 10.7 K in \citet{2011ApJ...742...27L}, its physical parameters were estimated by \citet{2018ApJS..235...39C} using synchrotron self-Compton (SSC)/Klein-Nishina model. \\
	\\
	During 2008 -- 2012, variability in $R$ band was present, but $B-R$ chromatism in the $\sim$1 magnitude range of $R$ band has not been revealed, in \citet{2015A&A...573A..69W}. Since 2011 the authors of \citet{2016A&A...587A.112T} investigated the long term periodicity in $V$ and $R$ bands using Lomb-Scargle method and CLEANEST algorithm \citep{1987AJ.....93..968R}. Period was discovered only in $R$ band of 432 days with Lomb-Scargle method, and 435.7 days with CLEANEST algorithm. In \citet{2018A&A...611A..52T} was detected period of 35 days of variability in optical $G$ band, for period of observations from 2013 to 2016. In June 2015 during three hours of monitoring, object did not show the variability in $V$ band, showed possible variability in $R$ band, and a strong RWB trend of the optical spectrum, \citet{2021ApJS..257...41K}. The authors of paper \citet{2016A&A...593A..98L} discovered correlation between optical $R$ band and radio light curves at 15 GHz. Brightness variability in data collected over 12 years in $X$-ray was analysed by \citet{2009ApJ...696.2170R}. The period of about one year was explained as observational artifact. \\
	\\
	With almost 2 magnitude changes in brightness this object has the highest brightness changes. In the $R$ band has the maximum brightness 14.371 mag on the date 28 August 2016 (this period was not covered with observations in $V$ band), the next maximum of 14.458 mag occurred on 20 July 2018, and was detected in $V$ band of 14.888 mag. The slope and Pearson's coefficient of colour--time is positive, but close to 0, with a probability close to 0.9. The slope of the colour--magnitude dependencies is around 0, Pearson's coefficient is positive. The colour has small tendencies to change during time of observations. From colour magnitude dependencies we can conclude that BWB variations are present, but we noted that during observational period two tendencies of colour variation in dependence of $R$ magnitude are present. One in the beginning of observations from 2013 to 2016, and the second from 2016 to the end of observational cycle. If we separate the data in two sections in first three years RWB tendencies, and in the next three years period BWB variations were present, see Table \ref{table:1722}, and Fig. \ref{fig:Fig1722_123}. In the second period were detected both minimum and maximum of object brightness. In the last 300 days brightness decreased by 1.6 magnitude, and reached the minimum 16.8, and 16.3 mag, in $V$, and $R$ band, respectively. 
	
	\begin{table}
		\caption{The colour – magnitude dependencies for 1722+119.}  
		\label{table:1722} 
		\centering 
		\begin{footnotesize}
			\begin{tabular}{c r r r c }  
				\hline\hline    
				Source & Slope & Intercept & r & P \\\hline   
				before 2016 & -0.051 $\pm$ 0.018 & 1.21 $\pm$ 0.28 & -0.32 & 0.1510\\
				after 2016 & 0.034 $\pm$ 0.006 & -0.08 $\pm$ 0.08 & 0.64 & 0.0012\\\hline      
				\multicolumn{5}{l}{Notes. r - Pearson's coefficient and P - null hypothesis probability.}
		\end{tabular}\end{footnotesize}
		
	\end{table}
	
	\subsection{1741+597} The source was catalogued in the same year in two papers \citet{1991ApJS...75.1011G}, and \citet{1991ApJS...75....1B}. In \citet{1998ApJS..118..127L} the source was classified as BL Lac. The redshift was determined by photometry $z=0.415$ by \citet{2009ApJS..180...67R}. The source is ISP according to papers \citet{2006A&A...445..441N}), and \citet{2011ApJ...743..171A}. We adopt $log \nu_{peak}= 15.20$ given in \citet{2017ApJ...841..113M} and classify source as HPS. With the name 9Y-MST J1742+5946 source is in the MST catalogue of $\gamma$-ray source candidates above 10 GeV \citep{2018A&A...619A..23C}. In \citet{2018ApJS..235...39C} are estimated physical parameters of jet using synchrotron self-Compton (SSC)/Thomson model. \\
	\\
	The host galaxy was detected by \citet{2003A&A...400...95N}, the $R$ band magnitude of nucleus (17.06$\pm$0.03) and host galaxy (19.33$\pm$0.06) with effective radius of 1.6$\pm$0.2 arcsec were presented in the paper. This source is the second one with respect to the brightness change, with about 1.6 mag. In the last 250 days object became brighter by 1.2 magnitude. The colour has tendencies to change during time of observations, these changes are about 0.3 mag. From colour magnitude dependencies we can conclude that BWB variations are present. 
	
	
	\section{Discussions and Conclusions}\label{sec:5}

	To understand the emission mechanism of blazars on diverse timescales, flux variability study play an important role and can provide the information about emitting region e.g. size, location and its dynamics \citep{2003A&A...400..487C}. Variability in blazars can be of intrinsic as well as of extrinsic nature. The extrinsic variability in blazars is caused by frequency-dependent interstellar scintillation and is found to be dominant mechanism in low frequency radio observations \citep{1995ARA&A..33..163W}. Intrinsic mechanism operates across whole EM spectrum, and include directly those causing variation in the jet emission. In blazars, the Doppler boosted non-thermal radiation from the jet dominates on the thermal emission from the accretion disk \citep[e.g.,][and references therein]{1993ApJ...406..420M,1993ApJ...411..602C,1995ARA&A..33..163W,1995PASP..107..803U,1997ARA&A..35..445U,2019ARA&A..57..467B}. On diverse timescales such as IDV, STV and LTV variability in blazars can be explained by various jet based models e.g. shock-in-jet, turbulence behind the shock, or other irregularities in the jet flow produced by variations in the outflow parameters \citep[e.g.,][and references therein]{1979ApJ...232...34B,1985ApJ...298..114M,2013A&A...558A..92B,2014ApJ...780...87M,2015JApA...36..255C}. Variation in the jet geometry due to changing jet direction may lead to the variations in the Doppler factor, and Lorentz factor of the relativistic blobs moving along the jet, which in turn can lead to STV and LTV in the blazar \citep{2009A&A...494..527H}. During the low flux states of blazars, the variability can be attributed to accretion disk instabilities since thermal radiation from the central region of blazars may dominate over jet emission \citep{1993ApJ...406..420M,1993ApJ...411..602C}. \\  
	\\
	The variation of Doppler factor can cause slight deviation in the optical spectra of the blazar from a power-law which leads to a BWB trend \citep{2006A&A...453..817V}. The increase in brightness of the blazar due to injection of fresh electrons with an energy distribution harder than that of the previously cooled ones can also cause BWB trend \citep{1998A&A...333..452K,2002PASA...19..138M}.
	A RWB trend indicates an increase of thermal contribution at the blue end of the spectrum, with decrease in nonthermal jet emission \citep{2006A&A...453..817V,2012AJ....143...23G}. The presence of both BWB and RWB trends in some blazars can be explained by superposition of both blue and red emission components where the redder one is attributed to the synchrotron radiation from the relativistic jet while the blue component could come from the thermal emission from the accretion disk.\\
	\\
	In this paper we analysed the multi-band optical photometric data of 12 blazars selected from a sample of 47 AGNs detected by \citet{2011A&A...526A.102B}. Among these 12 blazars: 6 are BL Lacs, 4 are FSRQs and 2 show dual nature of BL Lac / FSRQ. During 14 April 2013 to 8 August 2019, the optical photometric observations of these blazars were carried out in $V$ and $R$ passbands  using 8 telescopes located in 4 countries in Europe. In our $\sim$ 6 years of observations, most of the blazars have shown significant flux and colour variations on STV and LTV timescales, and the variability pattern in $V$ and $R$ bands found to be similar. On the LTV timescale, the minimum variation of $\sim$ 0.2 mag is found in the blazar 1556+335 while the maximum variation of $\sim$ 2.0 mag is found in two blazars, namely 1722+119 and 1741+597. Four BL Lacs, two FSRQs and one blazar with dual nature show BWB trend. RWB trend is displayed by one BL Lac and one blazar with dual nature. The BL Lac 1722+119 shows RWB trend in the first about three years of observations, and BWB trend in the next about three years of observations. These trends show that in our sample of blazars and their observations, the most commonly found trends e.g. BWB in BL Lacs and RWB in FSRQs \citep[e.g.,][and references therein]{2006A&A...450...39G,2012MNRAS.425.3002G,2017MNRAS.472..788G} were not always found. In future we plan observations of more densely sampled light curves for extended period of time for these as well as several other blazars to make a better conclusion on BWB and RWB trends of BL Lacs and FSRQs.\\
	
	\section*{Acknowledgements}
	
	This research was supported by the Ministry of Science, Technological Development and Innovation of the Republic of Serbia (contract No.~451-03-47/2023-01/200002). GD acknowledges the support through the project F-187 of the Serbian Academy of Sciences and Arts, and the observing and financial grant support from the Institute of Astronomy and Rozhen NAO BAS through the bilateral joint research project "Gaia Celestial Reference Frame (CRF) and fast variable astronomical objects" (2020-2022,head – G. Damljanovic). ACG is partially supported by Chinese Academy of Sciences (CAS) President’s International Fellowship Initiative (PIFI) (grant no. 2016VMB073).\\
	\\
	Funding for the Sloan Digital Sky Survey IV has been provided by the Alfred P.~Sloan Foundation, the U.S. Department of Energy Office of Science, and the Participating Institutions. SDSS-IV acknowledges support and resources from the Center for High-Performance Computing at the University of Utah. The SDSS web site is {\url{www.sdss.org}}.
	SDSS-IV is managed by the Astrophysical Research Consortium for the Participating Institutions of the SDSS Collaboration including the Brazilian Participation Group, the Carnegie Institution for Science, Carnegie Mellon University, the Chilean Participation Group, the French Participation Group, Harvard-Smithsonian Center for Astrophysics, Instituto de Astrof\'isica de Canarias, The Johns Hopkins University, Kavli Institute for the Physics and Mathematics of the Universe (IPMU) / University of Tokyo, the Korean Participation Group, Lawrence Berkeley National Laboratory, Leibniz Institut f\"ur Astrophysik Potsdam (AIP), Max-Planck-Institut f\"ur Astronomie (MPIA Heidelberg), Max-Planck-Institut f\"ur Astrophysik (MPA Garching), Max-Planck-Institut f\"ur Extraterrestrische Physik (MPE), National Astronomical Observatories of China, New Mexico State University, New York University, University of Notre Dame, Observat\'ario Nacional / MCTI, The Ohio State University, Pennsylvania State University, Shanghai Astronomical Observatory, United Kingdom Participation Group, Universidad Nacional Aut\'onoma de M\'exico, University of Arizona, University of Colorado Boulder, University of Oxford, University of Portsmouth, University of Utah, University of Virginia, University of Washington, University of Wisconsin, Vanderbilt University, and Yale University.
	
	\section*{Data Availability}

	The light curves presented in this paper will be published in the electronic version of the Journal and in the CDS Vizier service, with the form illustrated in Table \ref{table:example}.

	
	
	\bibliographystyle{mnras}
	\bibliography{ref} 


\end{document}